\documentclass[a4paper,12pt,dvips,oneside]{article}
\usepackage{booktabs,graphicx,overcite,xspace,amsmath,amssymb,longtable,multirow,algorithm,algorithmic}
\usepackage[scanall]{psfrag}
\usepackage[small,sc]{caption}
\usepackage[dvips]{color}
\renewcommand\baselinestretch{1.8}

\renewcommand{\ge}{\geqslant}

\renewcommand{\tilde}{\widetilde}
\newcommand{\dsum}{\displaystyle\sum}
\newcommand{\mWt}{\tilde{\mathcal{W}}}
\newcommand{\mPt}{\tilde{\mathcal{P}}}
\newcommand{\mEt}{\tilde{\mathcal{E}}}
\newcommand{\mSt}{\tilde{\mathcal{S}}}
\newcommand{\mW}{\mathcal{W}}
\newcommand{\mP}{\mathcal{P}}
\newcommand{\mE}{\mathcal{E}}

\newcommand{\mT}{\mathcal{T}}
\begin{document}
\def\cam{Cambridge University Press, Cambridge }
\def\dover{Dover Publications, New York }
\def\mit{The MIT Press, Cambridge, Massachusetts }
\def\wiley{John Wiley and Sons, New York }
\def\springer{Springer-Verlag, New York }
\def\elsevier{Elsevier, Amsterdam }
\def\aciee{Angew. Chem. Int. Ed. Engl. }
\def\ac{Acta. Crystallogr. }
\def\acp{Adv. Chem. Phys. }
\def\acr{Acc. Chem. Res. }
\def\ajp{Am. J.~Phys. }
\def\ap{Ann. Physik }
\def\apc{Adv. Prot. Chem. }
\def\arpc{Ann. Rev. Phys. Chem. }
\def\bi{Bioinf. }
\def\bj{Biophys. J. }
\def\cccc{Coll. Czech. Chem. Comm. }
\def\cpc{Comp. Phys. Comm. }
\def\crev{Chem. Rev. }
\def\el{Europhys. Lett. }
\def\ic{Inorg. Chem. }
\def\ijmpc{Int. J.~Mod. Phys. C }
\def\ijqc{Int. J.~Quant. Chem. }
\def\jcis{J.~Colloid Interface Sci. }
\def\jcsft{J.~Chem. Soc., Faraday Trans. }
\def\jacs{J.~Am. Chem. Soc. }
\def\jas{J.~Atmos. Sci. }
\def\jbc{J.~Biol. Chem. }
\def\jcc{J.~Comp. Chem. }
\def\jcp{J.~Chem. Phys. }
\def\jce{J.~Chem. Ed. }
\def\jcscc{J.~Chem. Soc., Chem. Commun. }
\def\jetp{J.~Exp. Theor. Phys. (Russia) }
\def\jmb{J.~Mol. Biol. }
\def\jmsp{J.~Mol. Spec. }
\def\jmst{J.~Mol. Struct. }
\def\jncs{J.~Non-Cryst. Solids }
\def\jpa{J.~Phys. A }
\def\jpc{J.~Phys. Chem. }
\def\jpca{J.~Phys. Chem. A }
\def\jpcb{J.~Phys. Chem. B }
\def\jpcm{J.~Phys. Condensed Matter. }
\def\jpcs{J.~Phys. Chem. Solids. }
\def\jpsj{J.~Phys. Soc. Jpn. }
\def\jrnist{J.~Res. Natl. Inst. Stand. Technol. }
\def\mg{Math. Gazette }
\def\nat{Nature }
\def\nsb{Nat. Struct. Biol.}
\def\Pa{Physica A }
\def\Pd{Physica D }
\def\pac{Pure. Appl. Chem. }
\def\pccp{Phys. Chem. Chem. Phys. }
\def\phys{Physics }
\def\pmb{Philos. Mag. B }
\def\ptrsa{Philos. T. Roy. Soc. A }
\def\ptrsb{Philos. T. Roy. Soc. B }
\def\pnasu{Proc. Natl. Acad. Sci. USA }
\def\pr{Phys. Rev. }
\def\prep{Phys. Reports }
\def\pra{Phys. Rev. A }
\def\prb{Phys. Rev. B }
\def\prc{Phys. Rev. C }
\def\prd{Phys. Rev. D }
\def\pre{Phys. Rev. E }
\def\prl{Phys. Rev. Lett. }
\def\prsa{Proc. R. Soc. A }
\def\psfg{Proteins: Struct., Func. and Gen. }
\def\sci{Science }
\def\spj{Sov. Phys. JETP }
\def\ss{Surf. Sci. }
\def\tca{Theor. Chim. Acta }
\def\tpa{Theor. Prob. and Appl. }
\def\zpb{Z. Phys. B. }
\def\zpc{Z. Phys. Chem. }
\def\zpd{Z. Phys. D }
\def\aciee{Angew. Chem. Int. Ed. Engl. }
\def\ac{Acta. Crystallogr. }
\def\acp{Adv. Chem. Phys. }
\def\acr{Acc. Chem. Res. }
\def\ajp{Am. J.~Phys. }
\def\ap{Adv. Phys. }
\def\arpc{Ann. Rev. Phys. Chem. }
\def\cccc{Coll. Czech. Chem. Comm. }
\def\cpl{Chem. Phys. Lett. }
\def\crev{Chem. Rev. }
\def\dalton{J.~Chem. Soc., Dalton Trans. }
\def\el{Europhys. Lett. }
\def\faraday{J.~Chem. Soc., Faraday Trans. }
\def\fartrans{J.~Chem. Soc., Faraday Trans. }
\def\fdisc{J.~Chem. Soc., Faraday Discuss. }
\def\ic{Inorg. Chem. }
\def\ijqc{Int. J.~Quant. Chem. }
\def\jcis{J.~Colloid Interface Sci. }
\def\jcsft{J.~Chem. Soc., Faraday Trans. }
\def\jacs{J.~Am. Chem. Soc. }
\def\jas{J.~Atmos. Sci. }
\def\jcc{J.~Comp. Chem. }
\def\jcp{J.~Chem. Phys. }
\def\jce{J.~Chem. Ed. }
\def\jcscc{J.~Chem. Soc., Chem. Commun. }
\def\jetp{J.~Exp. Theor. Phys. (Russia) }
\def\jmc{J.~Math. Chem. }
\def\jmp{J.~Math. Phys. }
\def\jmsp{J.~Mol. Spec. }
\def\jmst{J.~Mol. Structure }
\def\jncs{J.~Non-Cryst. Solids }
\def\jpc{J.~Phys. Chem. }
\def\jpcm{J.~Phys. Condensed Matter. }
\def\jpsj{J.~Phys. Soc. Jpn. }
\def\jsp{J.~Stat. Phys. }
\def\mg{Math. Gazette }
\def\mp{Mol. Phys. }
\def\mpns{Mol. Phys.}
\def\nat{Nature }
\def\pac{Pure. Appl. Chem. }
\def\phys{Physics }
\def\pla{Phys. Lett. A }
\def\plb{Phys. Lett. B }
\def\phm{Philos. Mag. }
\def\pmb{Philos. Mag. B }
\def\pnas{Proc.\ Natl.\ Acad.\ Sci.\  USA }
\def\pr{Phys. Rev. }
\def\pra{Phys. Rev. A }
\def\prb{Phys. Rev. B }
\def\prc{Phys. Rev. C }
\def\prd{Phys. Rev. D }
\def\pre{Phys. Rev. E }
\def\prl{Phys. Rev. Lett. }
\def\prsa{Proc. R. Soc. A }
\def\ss{Surf. Sci. }
\def\sci{Science }
\def\tca{Theor. Chim. Acta }
\def\zpc{Z. Phys. Chem. }
\def\zpd{Z. Phys. D }
\def\zfpd{Z. Phys. D }
\def\zpdamc{Z. Phys. D }
\def\aciee{Angew. Chem. Int. Ed. Engl. }
\def\ac{Acta. Crystallogr. }
\def\acp{Adv. Chem. Phys. }
\def\acr{Acc. Chem. Res. }
\def\ajp{Am. J.~Phys. }
\def\am{Adv. Mater. }
\def\apl{Appl. Phys. Lett. }
\def\arpc{Ann. Rev. Phys. Chem. }
\def\mrsb{Mater. Res. Soc. Bull. }
\def\cccc{Coll. Czech. Chem. Comm. }
\def\cj{Comput. J. }
\def\cp{Chem. Phys. }
\def\cpc{Comp. Phys. Comm. }
\def\cpl{Chem. Phys. Lett. }
\def\crev{Chem. Rev. }
\def\el{Europhys. Lett. }
\def\fd{Faraday Disc. }
\def\ic{Inorg. Chem. }
\def\ijmpc{Int. J.~Mod. Phys. C }
\def\ijqc{Int. J.~Quant. Chem. }
\def\jcis{J.~Colloid Interface Sci. }
\def\jcsft{J.~Chem. Soc., Faraday Trans. }
\def\jacs{J.~Am. Chem. Soc. }
\def\jap{J.~Appl. Phys. }
\def\jas{J.~Atmos. Sci. }
\def\jcc{J.~Comp. Chem. }
\def\jcp{J.~Chem. Phys. }
\def\jce{J.~Chem. Ed. }
\def\jcscc{J.~Chem. Soc., Chem. Commun. }
\def\jetp{J.~Exp. Theor. Phys. (Russia) }
\def\jmsp{J.~Mol. Spec. }
\def\jmst{J.~Mol. Structure }
\def\jncs{J.~Non-Cryst. Solids }
\def\jpa{J.~Phys. A }
\def\jpc{J.~Phys. Chem. }
\def\jpcssp{J.~Phys. C: Solid State Phys. }
\def\jpca{J.~Phys. Chem. A. }
\def\jpcb{J.~Phys. Chem. B. }
\def\jpcm{J.~Phys. Condensed Matter. }
\def\jpcs{J.~Phys. Chem. Solids. }
\def\jpsj{J.~Phys. Soc. Jpn. }
\def\jpfmp{J.~Phys. F, Metal Phys. }
\def\mg{Math. Gazette }
\def\msr{Mater. Sci. Rep. }
\def\nat{Nature }
\def\njc{New J.~Chem. }
\def\njp{New J.~Phys. }
\def\pac{Pure. Appl. Chem. }
\def\phys{Physics }
\def\pma{Philos. Mag. A }
\def\pmb{Philos. Mag. B }
\def\pml{Philos. Mag. Lett. }
\def\pnasu{Proc. Natl. Acad. Sci. USA }
\def\pr{Phys. Rev. }
\def\prep{Phys. Reports }
\def\pra{Phys. Rev. A }
\def\prb{Phys. Rev. B }
\def\prc{Phys. Rev. C }
\def\prd{Phys. Rev. D }
\def\pre{Phys. Rev. E }
\def\prl{Phys. Rev. Lett. }
\def\prsa{Proc. R. Soc. A }
\def\pss{Phys. State Solidi }
\def\pssb{Phys. State Solidi B }
\def\rmp{Rev. Mod. Phys. }
\def\rpp{Rep. Prog. Phys. }
\def\sci{Science }
\def\ss{Surf. Sci. }
\def\tca{Theor. Chim. Acta }
\def\tetra{Tetrahedron }
\def\tams{Trans. Am. Math. Soc. }
\def\zpb{Z. Phys. B. }
\def\zpc{Z. Phys. Chem. }
\def\zpd{Z. Phys. D }
\def\currentyear{2006}

\title{Graph Transformation Method for Calculating Waiting Times in Markov Chains}
\author{Semen A. Trygubenko\footnote{E-mail: sat39@cam.ac.uk}~~and 
David J. Wales\footnote{E-mail: dw34@cam.ac.uk} \\
{\it University Chemical Laboratories, Lensfield Road,} \\
{\it Cambridge CB2 1EW, UK} }

\maketitle
\begin{abstract}
We describe an exact approach for calculating transition probabilities 
and waiting times in finite-state discrete-time Markov processes.
All the states and the rules for transitions between them must be known in advance. 
We can then calculate averages over a given ensemble of paths for both additive and
multiplicative properties in a non-stochastic and non-iterative fashion.
In particular, we can calculate the mean first-passage time between arbitrary groups of
stationary points for discrete path sampling databases,
and hence extract phenomenological rate constants.
We present a number of examples to demonstrate the efficiency
and robustness of this approach.
\end{abstract}

\section{Introduction}
\label{sec:intro}

Stochastic processes are widely used to treat phenomena with random factors and noise.
Markov processes are an important class of stochastic processes
for which future transitions do not depend upon how the current state was reached.
Markov processes restricted to a discrete, finite, or countably infinite state space
are called Markov chains~\cite{bolchgmt98,GrimmettS05a,GrimmettS05b}.
Many interesting problems of chemical kinetics
concern the analysis of finite-state samples of otherwise infinite state space~\cite{Wales03}.

When analysing the kinetic databases obtained from discrete path
sampling (DPS) studies~\cite{Wales02} it can be difficult to extract the
phenomenological rate constants for processes that occur over very long time scales~\cite{Wales03}.
DPS databases are composed of local minima of the potential energy surface (PES)
and the transition states that connect them.
While minima correspond to mechanically stable structures, the transition states
specify how these structures interconvert and can be used to calculate the corresponding rates.
Whenever the potential energy barrier for the event of interest is
large in comparison with $k_B T$ the event becomes rare,
where $T$ is the temperature and $k_B$ is Boltzmann's constant.

The most important tools previously employed to extract kinetic information from a
DPS stationary point database are
the master equation~\cite{kampen81},
kinetic Monte Carlo~\cite{bortzkl75,fichthornw91}
(KMC) and matrix multiplication (MM) methods~\cite{Wales02}.
The system of linear master equations in its matrix formulation can be solved numerically
to yield the time evolution of the occupation probabilities
starting from an arbitrary initial distribution.
This approach works well only for small problems, as
the diagonalisation of the transition matrix, ${\bf P}$, scales as the
cube of the number of states~\cite{Wales03}.
In addition, numerical problems arise when
the magnitude of the eigenvalues corresponding to the slowest relaxation mode
approaches the precision of the zero eigenvalue corresponding to equilibrium~\cite{Miller99}.
The KMC approach is a stochastic technique that is commonly used to simulate the dynamics of various
physical and chemical systems, examples being formation of crystal structures~\cite{blockks04},
nanoparticle growth~\cite{mukherjeesz03} and diffusion~\cite{bulnespr98}.
The MM approach provides a way to sum contributions to phenomenological two-state
rate constants from pathways that contain progressively more steps.
It is based upon a steady-state approximation,
and provides the corresponding solution to the linear master equation~\cite{kampen81,kunz95}.
The MM approach has been used to analyse DPS databases in a number of systems
ranging from Lennard-Jones clusters~\cite{Wales04,Wales02} to biomolecules~\cite{evansw04,evansw03}.

Both the standard KMC and MM formulations 
provide rates at a computational cost that generally grows exponentially as the
temperature is decreased.
In this contribution we describe alternative methods that are deterministic and formally
exact, where the computational requirements are independent of the temperature and the
time scale on which the process of interest takes place.

\subsection{Graph Theory Representation of a Finite-state Markov Chain}
\label{sec:graph_repr}

To fully define a Markov chain it is necessary to specify
all the possible states of the system and the rules for transitions between them.
Graph theoretical representations of finite-state Markov chains are widely
used~\cite{bolchgmt98,ApaydinBGHL02,ApaydinGVBL02,SinghalSP04}.
Here we adopt a digraph~\cite{chartrand77,CormenLRS01} representation,
where nodes represent the states and edges represent the transitions
with non-zero probabilities.
The edge $e_{i,j}$ describes a transition from node $j$ to node $i$
and has a probability $P_{i,j}$ associated with it,
which is commonly known as a routing or branching probability.
A node can be connected to any number of other nodes.
Two nodes of a graph are adjacent if there is an edge between them~\cite{dads}.

For digraphs all connections of a node are classified as incoming or
outgoing. The total number of incoming connections is the in-degree of a node, while the
total number of outgoing connections is the out-degree. In a symmetric digraph 
the in-degree and out-degree are the same for every node~\cite{CormenLRS01}.
$AdjIn[i]$ is the set of indices of all nodes that are connected to node $i$ via incoming edges
that finish at node $i$. Similarly, $AdjOut[i]$ is the set of indices of 
all the nodes that are connected to node $i$ via outgoing edges from node $i$.
The degree of a graph is the maximum degree of all of its nodes.
The expectation value for the degree of an undirected graph 
is the average number of connections per node.

For any node $i$ the transition probabilities $P_{j,i}$ add up to unity,
\begin{equation}
	\sum_j P_{j,i} = 1,
\end{equation}
where the sum is over all $j \in AdjOut[i]$. Unless specified otherwise
all sums are taken over the set of indices of adjacent nodes or,
since the branching probability is zero for non-adjacent nodes, over the set of all nodes.

In a computer program
dense graphs are usually stored in the form of adjacency matrices~\cite{CormenLRS01}.
For sparse graphs~\cite{chartrand77} a more compact but less efficient adjacency-lists-based
data structure exists~\cite{CormenLRS01}.
To store a graph representation of a Markov chain, in addition to connectivity information
(available from the adjacency matrix), the branching probabilities must be stored. Hence for dense graphs
the most convenient approach is to store a transition probability matrix~\cite{bolchgmt98}
with transition probabilities for non-existent edges set to zero. For sparse graphs,
both the adjacency list and a list of corresponding branching probabilities must be stored.

\subsection{The Kinetic Monte Carlo Method}
\label{sec:kmc}

The KMC method can be used to generate a memoryless (Markovian)
random walk and hence a set of trajectories connecting
initial and final states in a DPS database. Many trajectories are necessary to collect 
appropriate statistics.
Examples of pathway averages that are usually obtained with KMC are the mean path length
and the mean first-passage time. Here the KMC trajectory length is the number of states
(local minima of the PES in the current context)
that the walker encounters before reaching the final state. The first-passage time is defined
as the time that elapses before the walker reaches the final state.
For a given KMC trajectory the first-passage time is calculated as the sum of the mean waiting times
in each of the states encountered.

\begin{figure}[ht]
\centerline{\includegraphics{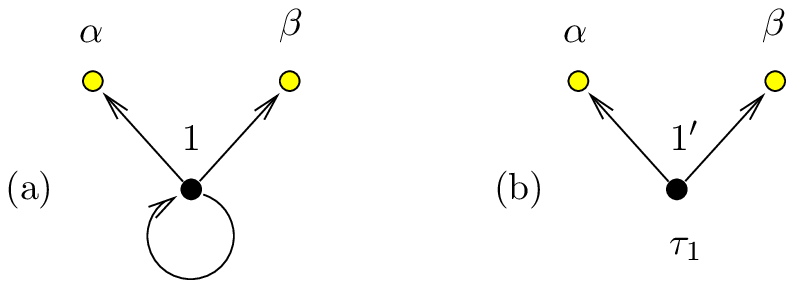}}
\caption{}
\label{fig:BKL}
\end{figure}

An efficient way to propagate KMC trajectories was suggested by
Bortz, Kalos, and Lebowitz~(BKL)~\cite{bortzkl75}.
According to the BKL algorithm, a step is chosen in such a way that the 
ratios between transition probabilities of different events are preserved,
but rejections are eliminated. Fig.~\ref{fig:BKL} explains this approach
for a simple discrete-time Markov chain.
The evolution of an ordinary KMC trajectory
is monitored by the `time' parameter $n\in\mathbb{W}$, which
actually corresponds to the number of steps~\cite{bolchgmt98}. The random walker is
in state $1$ at `time' $n=0$. The KMC trajectory is terminated whenever an absorbing state
is encountered. As $P_{1,1}$ approaches unity transitions out of state $1$ become rare.
To ensure that every time a random number is generated (one of the most time consuming steps in a KMC calculation)
a move is made to a neighbouring state we average over the transitions
from state $1$ to itself to obtain the Markov chain depicted in Fig.~\ref{fig:BKL}~(b).
Transitions from state $1$ to itself can be modelled by a Bernoulli process~\cite{wikipedia}
with the probability of success equal to $P_{1,1}$. The average `time' for escape from
state $1$ is obtained as
\begin{equation}
	\tau_1 = (1-P_{1,1})\dsum_{n=0}^{\infty} (n+1) (P_{1,1})^n = \dfrac{1}{(1-P_{1,1})},
\end{equation}
which can be used as a measure of the efficiency of trapping~\cite{Bar-haimK98}.
In the BKL scheme, transition probabilities out of state $1$ are renormalised as:
\begin{equation}
	P_{\alpha,1'} = \dfrac{P_{\alpha,1}}{1-P_{1,1}}, \qquad P_{\beta,1'} = \dfrac{P_{\beta ,1}}{1-P_{1,1}}.
\end{equation}
Similar ideas underlie the accelerated Monte Carlo algorithm suggested by Novotny~\cite{novotny94}.
Both the BKL and Novotny's methods
can be many orders of magnitude faster than the standard KMC method
when kinetic traps are present.

In chemical kinetics transitions out of a state are usually described
using a Poisson process, which can be considered a
continuous-time analogue of Bernoulli trials.
The transition probabilities are determined from the rates of the underlying transitions as
\begin{equation}
	P_{j,i} = \dfrac{k_{j,i}}{\displaystyle \sum_{\alpha} k_{\alpha,i}},
\end{equation}
where $k_{i,j}$ is the rate constant for transitions from $j$ to $i$, etc.
There may be several escape routes from a given state.
Transitions from any state to directly connected states are treated as
competing independent Poisson processes,
which together generate a new Poisson distribution~\cite{bulmer79}.
$n$ independent Poisson processes with rates
$k_1,\ k_2,\ k_3,\dots,\ k_n$ combine to produce a Poisson process with rate $k=\sum_{i=1}^{n} k_i$.
The waiting time for a transition to occur to any connected state
is then exponentially distributed as $k\exp(-kt)$~\cite{trumbo99}.

Given the exponential distribution of waiting times
the mean waiting time in state $i$ before escape, $\tau_i$, is $1/\sum_j k_{j,i}$,
and the variance of the waiting time is simply $\tau_i^2$.
When the average of the distribution of times is the property of interest,
and not the distribution itself, it is sufficient
to increment the simulation time by the mean waiting time rather than by
a value drawn from the appropriate distribution~\cite{Wales03,middleton03}.
This modification
to the original KMC formulation~\cite{reede81,voter05} reduces the cost of the method
and accelerates the convergence of KMC averages without affecting the results.

\subsection{Discrete Path Sampling}
\label{sec:dps}

The result of a DPS simulation is a database of local minima and transition
states from the PES~\cite{Wales02,Wales03,Wales04}.
To extract thermodynamic and kinetic properties from this database we
require partition functions for the individual minima and rate constants,
$k_{\alpha,\beta}$, for the elementary transitions between adjacent minima $\beta$ and $\alpha$.
We usually employ harmonic densities of states and statistical rate theory
to obtain these quantities, but these details are not important here.
To analyse the global kinetics we further assume Markovian transitions
between adjacent local minima, which produces a
set of linear (master) equations that 
governs the evolution of the occupation probabilities towards equilibrium~\cite{kampen81,kunz95}
\begin{equation}
	\dfrac{dP_\alpha(t)}{dt} = \sum_{\beta} k_{\alpha,\beta} P_\beta(t) - P_\alpha(t) \sum_{\beta} k_{\beta,\alpha},
	\label{eq:master_equation}
\end{equation}
where $P_\alpha(t)$ is the occupation probability of minimum $\alpha$ at time $t$.

All the minima are classified into sets $A$, $B$ and $I$,
where $A$ and $B$ are the two states of interest and $I$ corresponds to `intervening' minima.
When local equilibrium is assumed
within the $A$ and $B$ sets we can write
\begin{equation}
	P_a(t) = \dfrac{P_a^{\rm eq} P_A(t)}{P_A^{\rm eq}} \quad {\rm and} \quad 
	   P_b(t) = \dfrac{P_b^{\rm eq} P_B(t)}{P_B^{\rm eq}},
\end{equation}
where $P_A(t) = \sum_{a \in A} P_a(t)$ and $P_B(t) = \sum_{b \in B} P_b(t)$.
If the steady-state approximation is applied to all the 
intervening states $i \in I$, so that
\begin{equation}
	\dfrac{dP_i(t)}{dt} = 0,
\end{equation}
then Eq.~(\ref{eq:master_equation}) can be written as~\cite{Wales03} 
\begin{equation}
	\begin{array}{lll}
		\dfrac{dP_A(t)}{dt} &=& k_{A,B} P_B(t) - k_{B,A} P_A(t), \\
	\noalign{\medskip}
		\dfrac{dP_B(t)}{dt} &=& k_{B,A} P_A(t) - k_{A,B} P_B(t).
	\end{array}
\end{equation}
The rate constants $k_{A,B}$ and $k_{B,A}$
for forward and backward transitions between states $A$ and $B$
are the sums over all possible paths within the set of intervening minima of
the products of the branching probabilities corresponding to the elementary transitions for
each path:
\begin{equation}
	\begin{array}{lll}
	k_{A,B}^{\rm SS}
	&=& \displaystyle \sum_{a\leftarrow b}'
	\dfrac{\displaystyle k_{  a,i_1}}{\displaystyle \sum_{\alpha_1} k_{\alpha_1,i_1} }
	\dfrac{\displaystyle k_{i_1,i_2}}{\displaystyle \sum_{\alpha_2} k_{\alpha_2,i_2} }
	\cdots
	\dfrac{\displaystyle k_{i_{n-1},i_n}}{\displaystyle \sum_{\alpha_n} k_{\alpha_n,i_n}}
	\dfrac{\displaystyle k_{i_n,b}\, P_{b}^{\rm eq}}{\displaystyle P_B^{\rm eq}} \\
	\noalign{\medskip}
	&=&
	\displaystyle \sum_{a\leftarrow b}'
	P_{  a,i_1}
	P_{i_1,i_2}
	\cdots
	P_{i_{n-1},i_n}
	\dfrac{\displaystyle k_{i_n,b}\, P_{b}^{\rm eq}}{\displaystyle P_B^{\rm eq}}, 
	\label{eq:kABDPS}
	\end{array}
\end{equation}
and similarly for $k_{B,A}$~\cite{Wales02}.
The superscript `SS' specifies that the DPS rate constant formula
was derived employing the steady-state approximation, as in the previous versions of the DPS method~\cite{Wales02,Wales04}.
The sum
is over all paths that begin from a state $b \in B$ and end at a state $a \in A$, and
the prime indicates that paths are not allowed to revisit states in $B$.
In previous contributions~\cite{Wales02,Wales04,evansw04,evansw03}
this sum was evaluated using a weighted adjacency matrix multiplication method.
The contributions of individual discrete paths to the total rate constants were also
calculated in this way. However, analytic results from the theory of one-dimensional random
walk~\cite{Raykin92,MurthyK89,Trygubenko06} are now employed instead. It is also possible to evaluate rates without
invoking the steady-state approximation~\cite{TrygubenkoW06}, as discussed in the following sections.

\subsection{KMC and Steady-state Averages}
\label{sec:kmcanddps}

We now show that the evaluation of the steady-state sum in Eq.~(\ref{eq:kABDPS}) and the calculation of KMC
averages are two closely related problems.

For KMC simulations we define sources and sinks that
coincide with the set of initial states, $B$, and final states, $A$, respectively.
Every cycle of KMC simulation involves the generation of a single KMC trajectory
connecting a node $b \in B$ and a node $a \in A$. A source node $b$ is
chosen from set $B$ with probability $P_b^{\rm eq}/P_B^{\rm eq}$. 

We can formulate the calculation of the mean first-passage time from $B$ to $A$
in graph theoretical terms as follows. Let the digraph consisting of nodes for all local minima
and edges for each transition state be $\mathcal{G}$.
The digraph consisting of all nodes except those belonging to region $A$
is denoted by $G$.
We assume that there are no isolated nodes in $\mathcal{G}$,
so that all the nodes in $A$ can be reached from every node in $G$, in one or more steps.
Suppose we start a KMC simulation from a particular node $\beta\in G$. 
Let $P_\alpha(n)$ be the expected occupation probability of node $\alpha$
after $n$ KMC steps,
with initial conditions
$P_\beta(0)=1$ and $P_{\alpha\not=\beta}(0)=0$.
We further define an escape probability for each $\alpha\in G$ as the sum
of branching probabilities to nodes in $A$, i.e.
\begin{equation}
	\mathcal{E}_{\alpha}^{G} = \sum_{a\in A} P_{a,\alpha}.
\end{equation}
KMC trajectories terminate when they arrive at an $A$ minimum, and the
expected probability transfer to the $A$ region at the $n$th KMC step is
$\sum_{\alpha\in G}\mathcal{E}_{\alpha}^{G}P_\alpha(n)$.
If there is at least one escape route from $G$ to $A$ with a non-zero 
branching probability, then eventually all the occupation probabilities
in $G$ must tend to zero and
\begin{equation}
\Sigma_{\beta}^{G} = \sum_{n=0}^\infty \sum_{\alpha\in G}\mathcal{E}_{\alpha}^{G}P_\alpha(n) = 1
\end{equation}
for any $\beta\in G$.
We now rewrite $P_\alpha(n)$ as a sum over all $n$-step paths that start from
$\beta$ and end at $\alpha$ without leaving $G$. Each path contributes
to $P_\alpha(n)$ according to the appropriate product of $n$ branching probabilities, so
that
\begin{equation}
	\begin{array}{lll}
	\Sigma_{\beta}^{G}
	&=& \displaystyle \sum_{\alpha\in G} \mathcal{E}_{\alpha}^{G} \sum_{n=0}^\infty P_\alpha(n) \\
	\noalign{\medskip}
	&=& \displaystyle \sum_{\alpha\in G} \mathcal{E}_{\alpha}^{G} \sum_{n=0}^\infty \, \sum_{\Xi(n)}
	    P_{\alpha,i_{n-1}} P_{i_{n-1},i_{n-2}}\cdots P_{i_2,i_1} P_{i_1,\beta}  \\
	\noalign{\medskip}
	&=& \displaystyle \sum_{\alpha\in G} \mathcal{E}_{\alpha}^{G} \mathcal{S}_{\alpha,\beta}^{G} = 1,
	\end{array}
\label{eq:defS}
\end{equation}
where $\Xi(n)$ denotes the set of $n$-step paths
that start from $\beta$ and end at $\alpha$ without leaving $G$, and
the last line defines the pathway sum $\mathcal{S}_{\alpha,\beta}^{G}$.

It is clear from the last line of Eq.~(\ref{eq:defS}) that for fixed $\beta$ the
quantities $\mathcal{E}_{\alpha}^{G} \mathcal{S}_{\alpha,\beta}^{G}$ define a
probability distribution.
However, the pathway sums $\mathcal{S}_{\alpha,\beta}^{G}$ are not probabilities,
and may be greater than unity. 
In particular, $\mathcal{S}_{\beta,\beta}^{G}\ge1$ because the path of zero length
is included, which contributes one to the sum.
Furthermore, the normalisation condition on the last line of Eq.~(\ref{eq:defS})
places no restriction on $\mathcal{S}_{\alpha,\beta}^{G}$ terms for which 
$\mathcal{E}_{\alpha}^{G}$ vanishes.

We can also define a probability distribution for individual pathways.
Let $\mW_\xi$ be the product of branching probabilities associated
with a path $\xi$ so that
\begin{equation}
	\mathcal{S}_{\alpha,\beta}^{G} = \sum_{n=0}^\infty \sum_{\xi\in\Xi(n)} \mW_\xi 
	\equiv \sum_{\xi\in\alpha\leftarrow\beta} \mW_{\xi},
\end{equation}
where $\alpha\leftarrow\beta$ is the set of all appropriate paths from
$\beta$ to $\alpha$ of any length that can visit and revisit any node in $G$.
If we focus upon paths starting from minima in region $B$
\begin{equation}
\sum_{b\in B} \frac{P_b^{\rm eq}}{P_B^{\rm eq}}
\sum_{\alpha\in G} \mathcal{E}_{\alpha}^{G}  \sum_{\xi\in\alpha\leftarrow b} \mW_{\xi} =
\sum_{b\in B} \frac{P_b^{\rm eq}}{P_B^{\rm eq}}
\sum_{\alpha\in G_A} \mathcal{E}_{\alpha}^{G}  \sum_{\xi\in\alpha\leftarrow b} \mW_{\xi} = 1,
\end{equation}
where $G_A$ is the set of nodes in $G$ that are adjacent to $A$ minima in the
complete graph $\mathcal{G}$, since $\mathcal{E}_{\alpha}^{G}$ vanishes for all other nodes.
We can rewrite this sum as
\begin{equation}
\sum_{\xi\in G_A\leftarrow B} \frac{P_b^{\rm eq}}{P_B^{\rm eq}} \mathcal{E}_{\alpha}^{G} \mW_{\xi} =
\sum_{\xi\in A\leftarrow B} \frac{P_b^{\rm eq}}{P_B^{\rm eq}} \mW_{\xi} =
\sum_{\xi\in A\leftarrow B} \mathcal{P}_\xi =1,
\end{equation}
which defines the non-zero pathway probabilities $\mP_\xi$
for all paths that start from any node in set $B$ and finish at any node in set $A$.
The paths $\xi\in A\leftarrow B$ can revisit any minima in the $G$ set, but include
just one $A$ minimum at the terminus. Note that $\mW_\xi$ and $\mP_{\xi}$ can be used
interchangeably if there is only one state in set $B$.

The average of some property, $Q_{\xi}$, defined for each KMC trajectory, $\xi$,
can be calculated from the $\mathcal{P}_\xi$ as
\begin{equation}
	\left<Q_{\xi}\right>=\sum_{\xi\in A\leftarrow B} \mathcal{P}_{\xi}Q_{\xi}.
\end{equation}
Of course, KMC simulations avoid this complete enumeration by generating
trajectories with probabilities proportional to $\mathcal{P}_{\xi}$, so that a simple
running average can be used to calculate $\left<Q_{\xi}\right>$.
In the following sections we will develop alternative approaches based upon
evaluating the complete sum, which become increasingly efficient at low temperature.
We emphasise that these methods are only applicable to problems
with a finite number of states, which are assumed to be known in advance.

The evaluation of the sum based on the steady-state approximation defined in Eq.~(\ref{eq:kABDPS})
can also be rewritten in terms of pathway probabilities:
\begin{equation}
	\begin{array}{lll}
	k_{A,B}^{\rm DPS}
	&=& \displaystyle \sum_{n=0}^\infty \sum_{\Xi(n)}' P_{a,i_1} P_{i_1,i_2}
		\cdots P_{i_{n-1},i_n}
		\dfrac{\displaystyle k_{i_n,b}\, P_{b}^{\rm eq}}{\displaystyle P_B^{\rm eq}}, \\
	\noalign{\medskip}
	&=& \displaystyle \sum_{n=0}^\infty \sum_{\Xi(n)}' P_{a,i_1} P_{i_1,i_2}
		\cdots P_{i_{n-1},i_n} P_{i_n,b} \tau_b^{-1}
		\dfrac{\displaystyle P_{b}^{\rm eq}}{\displaystyle P_B^{\rm eq}} \\
	\noalign{\medskip}
	&=& \displaystyle \sum_{\xi\in A\leftarrow B}' \mathcal{P}_\xi \tau_b^{-1},
	\end{array}
	\label{eq:kABDPS2}
\end{equation}
where the prime on the summation indicates that the paths are not allowed to revisit
the $B$ region. We have also used the fact that $k_{i_n,b}=P_{i_n,b}/\tau_b$.

A digraph representation of the 
restricted set of pathways defined in Eq.~(\ref{eq:kABDPS2})
can be created if we allow sets of sources and sinks to overlap.
In that case all the nodes $A \cup B$ are defined to be sinks
and all the nodes in $B$ are sources, i.e.~every node in
set $B$ is both a source and a sink.
The required sum then includes all the pathways that finish at sinks of type $A$, but not
those that finish at sinks of type $B$.
This situation, where sets of sources and sinks (partially) overlap,
is discussed in detail in \S\ref{sec:STO}.

\subsection{Mean Escape Times}
\label{sec:SigmaTau}

Since the mean first-passage time between states $B$ and $A$, or the escape time
from a subgraph, is of particular interest, we first illustrate a means to
derive formulae for these quantities in terms of pathway probabilities.

The average time taken to traverse 
a path $\xi = \alpha_1,\alpha_2,\alpha_3,\dots,\alpha_{l(\xi)}$ is calculated as
$\mathfrak{t}_{\xi} = \tau_{\alpha_1}+\tau_{\alpha_2}+\tau_{\alpha_3},\dots,\tau_{\alpha_{l(\xi)-1}}$, where
$\tau_\alpha$ is the mean waiting time for escape from node $\alpha$, as above,
$\alpha_k$ identifies the $k$th node along path $\xi$,
and $l(\xi)$ is the length of path $\xi$.
The mean escape time from a graph $G$ if started from node $\beta$ is then
\begin{equation}
	\mT^{G}_\beta = \sum_{\xi \in A\leftarrow\beta} \mathcal{P}_{\xi} \mathfrak{t}_{\xi}.
	\label{eq:tauCN}
\end{equation}
If we multiply every branching probability, $P_{\alpha,\beta}$, that appears in $\mathcal{P}_{\xi}$
by $\exp(\zeta \tau_\beta)$ then the mean escape time can be obtained as:
\begin{equation}
	\begin{array}{lll}
	\mT^{G}_\beta
	&=& \displaystyle \left[ \frac{d}{d\zeta} \left( \sum_{\xi \in A\leftarrow\beta}
		P_{\alpha_{l(\xi)},\alpha_{l(\xi)-1}} e^{\zeta\tau_{l(\xi)-1}} 
		P_{\alpha_{l(\xi)-1},\alpha_{l(\xi)-2}} e^{\zeta\tau_{l(\xi)-2}}  \ldots
		P_{\alpha_2,\alpha_1} e^{\zeta\tau_{\alpha_1}}  
		\right) \right]_{\zeta=0} \\
	\noalign{\medskip}
	&=& \displaystyle \left[ \frac{d}{d\zeta} \left( \sum_{\xi \in A\leftarrow\beta}
		P_{\alpha_{l(\xi)},\alpha_{l(\xi)-1}} 
		P_{\alpha_{l(\xi)-1},\alpha_{l(\xi)-2}} \ldots
		P_{\alpha_2,\alpha_1} e^{\zeta\mathfrak{t}_\xi}  
		\right) \right]_{\zeta=0} \\
	\noalign{\medskip}
	&=& \displaystyle \sum_{\xi \in A\leftarrow\beta} \mathcal{P}_{\xi} \mathfrak{t}_{\xi}.
	\end{array}
	\label{eq:tauCN2}
\end{equation}
This approach is useful if we have analytic results for
the total probability $\Sigma^{G}_\beta$, which may then be manipulated into
formulae for $\mT^{G}_\beta$, and is standard practice in probability theory~\cite{GoldhirschG86,GoldhirschG87}.
The quantity $P_{\alpha,\beta}e^{\zeta\tau_\beta}$ is similar to the `$\zeta$~probability' described in Ref.~\citen{GoldhirschG86}.
Analogous techniques are usually employed to obtain $\mT^{G}_\beta$
and higher moments of the first-passage time distribution
from analytic expressions for the first-passage probability generating function
(see, for example, Refs.~\citen{Raykin92,MurthyK89}).
We now define $\tilde{P}_{\alpha,\beta} = P_{\alpha,\beta} e^{\zeta\tau_\beta}$ and the
related quantities
\begin{equation}
	\begin{array}{rll}
	\mEt_{\alpha}^{G}
	&=& \dsum_{a\in A} \tilde{P}_{a,\alpha}
	 = \mE_{\alpha}^{G} e^{\zeta\tau_\alpha}, \\
\noalign{\medskip}
	\mWt_\xi
	&=& \tilde{P}_{\alpha_{l(\xi)},\alpha_{l(\xi)-1}} \tilde{P}_{\alpha_{l(\xi)-1},\alpha_{l(\xi)-2}} \ldots
	\tilde{P}_{\alpha_2,\alpha_1}
	 = \mW_\xi e^{\zeta\mathfrak{t}_\xi},  \\
\noalign{\medskip}
	\mPt_\xi
	&=& \tilde{\mathcal{W}}_\xi P_b^{\rm eq}/P_B^{\rm eq}, \\
\noalign{\medskip}
	\mSt_{\alpha,\beta}^{G}
	&=& \dsum_{\xi\in \alpha\leftarrow\beta} \mWt_{\xi}, \\
\noalign{\medskip}
	{\rm and} \quad \tilde{\Sigma}_\beta^{G}
	&=& \dsum_{\alpha\in G} \mEt_{\alpha}^{G} \mSt_{\alpha,\beta}^{G}. 
	\end{array}
\label{eq:SigmaG}
\end{equation}
Note that $\left[\mEt_{\alpha}^{G}\right]_{\zeta=0} = \mE_{\alpha}^{G}$ etc.,
while the mean escape time can now be written as
\begin{equation}
	\mT^{G}_\beta = \left[ \frac{d\tilde{\Sigma}_\beta^{G}}{d\zeta} \right]_{\zeta=0}.
	\label{eq:TauG}
\end{equation}
In the remaining sections we show how to calculate the pathway probabilities,
$\mathcal{P}_\xi$, exactly, along with pathway averages, such as the waiting time.

\section{Complete graphs}
\label{sec:complete_graphs}

In a complete digraph each pair of nodes is connected by
two oppositely directed edges.\cite{weisstein04cg} The complete graph with $N$ graph nodes is denoted
$K_N$, and has $N$ nodes and $N(N-1)$ edges,
remembering that we have two edges per connection.
Due to the complete connectivity we need only consider two cases: 
when the starting and finishing nodes are the same and when they are distinct.
We employ complete graphs for the purposes of generality. An arbitrary
graph $G_N$ is a subgraph of $K_N$ with transition probabilities for non-existent
edges set to zero. All the results in this section
are therefore equally applicable to arbitrary graphs.

The $\mathcal{S}_{\alpha,\beta}^{K_3}$ can be derived analytically:
\begin{equation}
	\label{eq:PK3}
	\begin{array}{lll}
	\mathcal{S}_{1,1}^{K_3} &=& \displaystyle\sum_{n=0}^{\infty} \left(
	\left( \mW_{1,2,1}+\mW_{1,3,1}+\mW_{1,2,3,1}+\mW_{1,3,2,1} \right)
	\sum_{m=0}^{\infty} \left(\mW_{2,3,2} \right)^m\right)^n \\
	\noalign{\medskip}
	&=& \dfrac{1-\mW_{2,3,2}}{1-\mW_{1,2,1}-\mW_{2,3,2}-\mW_{1,3,1}
	-\mW_{1,2,3,1}-\mW_{1,3,2,1}},\\
	\noalign{\bigskip}
	\mathcal{S}_{2,1}^{K_3} &=& \displaystyle\sum_{n=0}^{\infty} \left(\mW_{2,3,2}\right)^n
	\left( P_{2,1} + \mW_{2,3,1} \right) \mathcal{S}_{1,1}^{K_3} \\
	\noalign{\medskip}
	&=& \dfrac{P_{2,1}+\mW_{2,3,1}}{1-\mW_{1,2,1}-\mW_{2,3,2}-\mW_{1,3,1}
	-\mW_{1,2,3,1}-\mW_{1,3,2,1}},
	\end{array}
\end{equation}
where, as before, $\mW_{1,2,1}=P_{1,2}P_{2,1}$, etc.
The results for any other possibility can be obtained by permuting the node indices appropriately.

Pathway sums for larger complete graphs can be obtained by recursion.
For $\mathcal{S}_{N,N}^{K_N}$ any path leaving from and returning to $N$ can
be broken down into a step out of $N$ to any $i<N$, all possible paths between
$i$ and $j<N-1$ within $K_{N-1}$, and finally a step back to $N$ from $j$.
All such paths can be combined together in any order, so we have a multinomial
distribution:\cite{goldberg60}
\begin{equation}
	\begin{array}{lll}
	\mathcal{S}_{N,N}^{K_N}
	&=& \displaystyle \sum_{n=0}^{\infty} \left(
		\sum_{i=1}^{N-1} \left(
			\sum_{j=1}^{N-1} \left(
				P_{N,j}\mathcal{S}_{j,i}^{K_{N-1}}P_{i,N}
			\right)
		\right)
	\right)^n \\
	&=& \displaystyle \left(1-\sum_{i=1}^{N-1} 
         \sum_{j=1}^{N-1} P_{N,j}\mathcal{S}_{j,i}^{K_{N-1}}P_{i,N}\right)^{-1}. 
	\label{eq:PKN1}
	\end{array}
\end{equation}
To evaluate $\mathcal{S}_{1,N}^{K_N}$ we break down the sum into all paths that depart
from and return to $N$, followed by all paths that leave node $N$ and reach node 1 without returning to $N$.
The first contribution corresponds to a factor of $\mathcal{S}_{N,N}^{K_N}$, and the second 
produces a factor $P_{i,N}\mathcal{S}_{1,i}^{K_{N-1}}$ for every $i<N$:
\begin{equation}
\mathcal{S}_{1,N}^{K_N} = \mathcal{S}_{N,N}^{K_N}\sum_{i=1}^{N-1} \mathcal{S}_{1,i}^{K_{N-1}}P_{i,N},
\label{eq:PKN2}
\end{equation}
where $\mathcal{S}_{1,1}^{K_1}$ is defined to be unity. 
Any other $\mathcal{S}_{\alpha,\beta}^{K_N}$ can be obtained by a permutation of node labels.

Algorithm~\ref{alg:ProbForKn} provides an implementation of the above formulae
optimised for incomplete graphs.
The running time of Algorithm~\ref{alg:ProbForKn} depends strongly on the graph density.
(A digraph in which the number of edges is close to the maximum value of $N(N-1)$
is termed a dense digraph \cite{dads}.)
For $K_N$ the algorithm runs in $\mathcal{O}(N^{2N})$ time, while for an
arbitrary graph it scales as $\mathcal{O}(d^{2N})$, where $d$ is the average degree of
the nodes. 
For chain graphs the algorithm runs in $\mathcal{O}(N)$ time and has
linear memory requirements.
For complete graphs an alternative implementation
with $\mathcal{O}((N!)^2)$ scaling is also possible.

Although the scaling of the above algorithm with $N$ may appear disastrous,
it does in fact run faster than standard KMC and MM approaches
for graphs where the escape probabilities
are several orders of magnitude smaller than the
transition probabilities (Algorithm~\ref{alg:ProbForKn}).
Otherwise, for anything but moderately branched chain graphs, Algorithm~\ref{alg:ProbForKn} is
significantly more expensive.
However, the graph-transformation-based method presented in \S\ref{sec:tauKN} yields both
the pathway sums and the mean escape times for a complete graph $K_N$
in $\mathcal{O}(N^3)$ time, and is the fastest approach that we have found.

\begin{figure}[ht]
\centerline{\includegraphics{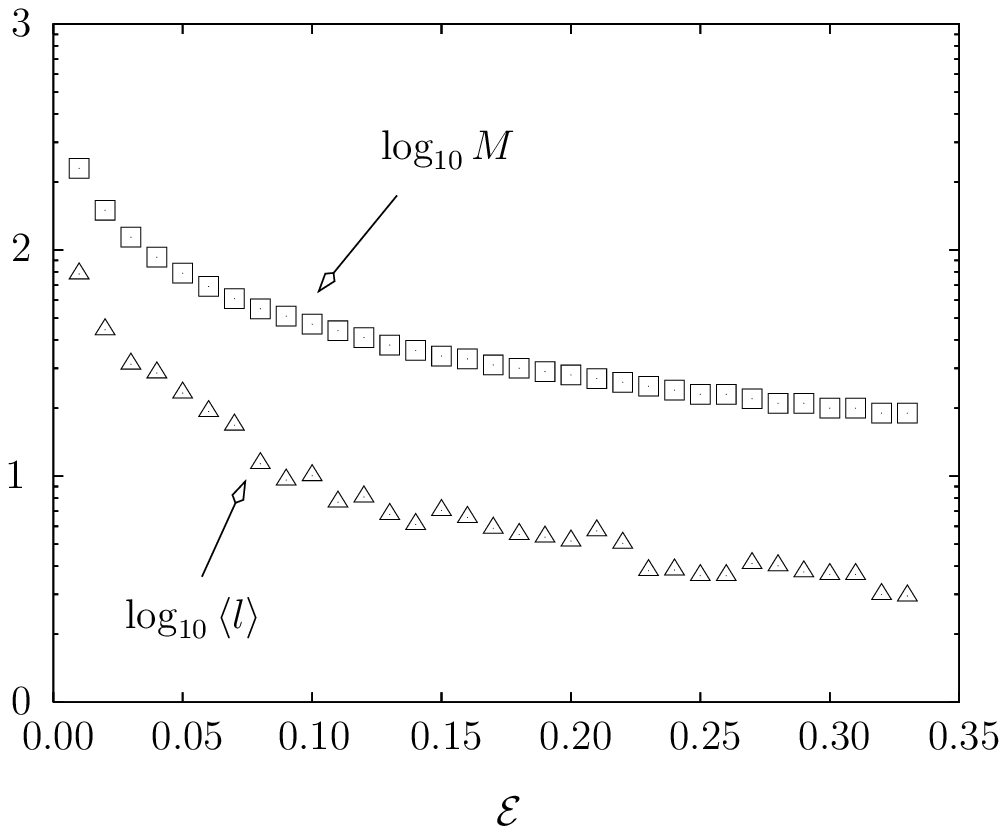}}
\caption{}
\label{fig:K3_computational_cost}
\end{figure}

Mean escape times for $K_3$ are readily obtained from the results in
equation (\ref{eq:PK3}) by applying the method outlined in \S\ref{sec:SigmaTau}:
\begin{equation}
\mT^{K_3}_1 = \dfrac{
	\tau_1 (1-\mW_{2,3,2}	 )
   + \tau_2 (P_{2,1}+\mW_{2,3,1}) 
   + \tau_3 (P_{3,1}+\mW_{3,2,1}) 
}{ 1-\mW_{1,2,1}-\mW_{2,3,2}-\mW_{3,1,3}-\mW_{1,2,3,1}-\mW_{1,3,2,1} }.
\label{eq:tauK3}
\end{equation}
We have verified this result numerically for various values of the
parameters $\tau_i$ and $P_{\alpha,\beta}$
and obtained quantitative agreement.
Fig.~\ref{fig:K3_computational_cost} demonstrates
how the advantage of exact summation over KMC and MM
becomes more pronounced as the escape probabilities become smaller.

\subsection{Mean escape time from $K_N$}
\label{sec:tauKN}

The problem of calculating the properties of a random walk on irregular networks was addressed previously
by Goldhirsch and Gefen~\cite{GoldhirschG86,GoldhirschG87}. They described a generating-function-based method
where an ensemble of pathways is partitioned into `basic walks'.
To the best of our knowledge only one~\cite{KahngR89} out of the 30 papers~\cite{
ZhengLW95,KimL95,BressloffDK96,RevathiBLM96,KimCK98,KimKK00,AsikainenHA02,PuryC03,SlutskyKM04,SlutskyM04,
MurthyK89,Ben-AvrahamRC89,GefenG89,MatanH89,HauckeWBUW90,RevathiB93a,Raykin92,NoskowiczG90,RevathiB93b,BalakrishnanV95,
GoldhirschG87,KersteinP87,GefenG87,GefenG85,HausK87,NoskowiczG87,LandauerB87,Tao87,KoplikRW88,KahngR89}
that cite the work of Goldhirsch and Gefen~\cite{GoldhirschG86} is an application, perhaps due to the complexity of the method.
Here we present a graph transformation (GT) approach
for calculating the pathway sums and the mean escape times for $K_N$.
In general, it is applicable to arbitrary digraphs, but
the best performance is achieved when the graph in question is dense.
A sparse-optimised version of the GT method will be discussed in \S\ref{sec:RN}.

The GT approach is similar in spirit to the ideas that lie behind the mean value analysis and
aggregation/disaggregation techniques commonly used in the performance and reliability evaluation of
queueing networks.\cite{bolchgmt98,dijk93,gelenbep98,conwayg89}
It is also loosely related to dynamic graph
algorithms \cite{eppsteingi97,cherkasskygr94,ramalingamr96a,ramalingamr96b}, which are
used when a property is calculated on a graph subject to dynamic changes,
such as deletions and insertions of nodes and edges.
The main idea is to progressively remove nodes from a graph
whilst leaving the average properties of interest unchanged.
For example, suppose we wish to remove node $x$ from graph $G$ to
obtain a new graph $G'$. Here we assume that $x$ is neither source nor sink.
Before node $x$ can be removed the property of interest is
averaged over all the pathways that include the edges between nodes $x$
and $i \in Adj[x]$. 
The averaging is performed separately for every node $i$.
We will use the waiting time as an example of such a property and
show that the mean first-passage time in the original and transformed graphs is the same.

The theory is an extension of the results used to perform jumps
to second neighbours in previous KMC simulations.\cite{Wales02}
Consider KMC trajectories that arrive at node $i$, which is adjacent to $x$.
We wish to step directly from $i$ to any vertex in the set of nodes
$\Gamma$ that are adjacent to $i$ or $x$, excluding these two nodes themselves.
To ensure that the mean first-passage times from
sources to sinks calculated in $G$ and $G'$ are the same we must
define new branching probabilities, $P_{\gamma,i}'$ from $i$ to all $\gamma\in\Gamma$,
along with a new waiting time for escape from $i$, $\tau_i'$.
Here, $\tau_i'$ corresponds to the mean waiting time for escape from
$i$ to any $\gamma\in\Gamma$, while the modified branching probabilities
subsume all the possible recrossings involving node $x$ that could occur
in $G$ before a transition to a node in $\Gamma$.
Hence the new branching probabilities are~\cite{TrygubenkoW06}:
\begin{equation}
P_{\gamma,i}' = (P_{\gamma,x}P_{x,i}+P_{\gamma,i}) 
\sum_{m=0}^\infty (P_{i,x}P_{x,i})^m = (P_{\gamma,x}P_{x,i}+P_{\gamma,i})/(1-P_{i,x}P_{x,i}).
\end{equation}
This formula can also be applied if either $P_{\gamma,i}$ or $P_{\gamma,x}$ vanishes.

It is easy to show that the new branching probabilities are normalised:
\begin{equation}
\sum_{\gamma\in\Gamma} \frac{P_{\gamma,x}P_{x,i}+P_{\gamma,i}}{1-P_{i,x}P_{x,i}}
= \frac{(1-P_{i,x})P_{x,i}+(1-P_{x,i})}{1-P_{i,x}P_{x,i}}
= 1.
\end{equation}
To calculate $\tau_i'$ we use the method of \S\ref{sec:kmcanddps}:
\begin{equation}
\tau_i' = \left[ \frac{d}{d\zeta}
\sum_{\gamma\in\Gamma}\frac{P_{\gamma,x}P_{x,i}e^{\zeta(\tau_x+\tau_i)}
+P_{\gamma,i}e^{\zeta\tau_i}}{1-P_{i,x}P_{x,i}e^{\zeta(\tau_x+\tau_i)}} \right]_{\zeta=0}
= \frac{\tau_i+P_{x,i}\tau_x}{1-P_{i,x}P_{x,i}}.
\end{equation}
The modified branching probabilities and waiting times could be used in a KMC
simulation based upon graph $G'$. 
Here we continue to use the notation of \S\ref{sec:kmcanddps}, where sinks
correspond to nodes $a\in A$ and sources to nodes in $b\in B$, and $G$ contains 
all the nodes in $\mathcal{G}$ expect for the $A$ set,
as before.
Since the modified branching probabilities, $P_{\gamma,i}'$, in $G'$ subsume the sums over all
paths from $i$ to $\gamma$ that involve $x$ we would expect the sink probability,
$\Sigma_{a,b} = \sum_{\xi\in a\leftarrow b} \mW_\xi$, of
a trajectories starting at $b$ ending at sink $a$, to be conserved.
However, since each trajectory exiting from $\gamma\in\Gamma$ acquires a time increment equal to
the average value, $\tau_i'$, the mean first-passage times to individual sinks,
$\mT_{a,b} = \frac{d}{d\zeta}\widetilde{\Sigma}_{a,b}$, are not conserved in $G'$ (unless there is a single sink). 
Nevertheless, the overall mean first-passage time to $A$ is
conserved, i.e.~$\mT_{b}^{G'}=\mT_{b}^{G}$.
To prove these results formally
consider the effect of removing node $x$ on trajectories reaching node $i\in Adj[x]$
from a source node $b$. The sink probability for a particular sink $a$ is
\begin{equation}
	\begin{array}{lll}
	\Sigma_{a,b}
	&=&\dsum_{\xi\in a\leftarrow b} \mW_\xi \\
	&=&\dsum_{\xi_3\in \Xi'} \mW_{\xi_3} +
        \dsum_{\xi_1\in i\leftarrow b} \mW_{\xi_1}
	   \dsum_{\gamma\in\Gamma} (P_{\gamma,i}+P_{\gamma,x}P_{x,i})
	   \dsum_{m=0}^\infty(P_{i,x}P_{x,i})^m
	   \dsum_{\xi_2\in a\leftarrow\gamma} \mW_{\xi_2} \\
	&=&\dsum_{\xi_3\in \Xi'} \mW_{\xi_3} +
        \dsum_{\xi_1 \in i\leftarrow b} \mW_{\xi_1}
	   \dsum_{\gamma\in\Gamma} P_{\gamma,i}'
	   \dsum_{\xi_2\in a\leftarrow\gamma} \mW_{\xi_2},
    \end{array}
\end{equation}
and similarly for any other node adjacent to $x$
and any other pathway that revisits $i$ other than
by immediate recrossings of the type $i\rightarrow x\rightarrow i$.
$\Xi'$ is the ensemble of all paths for which probabilistic weights
cannot be written in the form defined by the last term of the above equation.
Hence the transformation preserves the individual sink probabilities for any source.

Now consider how removing node $x$ from each trajectory not included in $\Xi'$
affects the mean first-passage time, $\mT_{a,b}$,
using the approach of \S\ref{sec:kmcanddps}.
\begin{equation}
	\begin{array}{lll}
&&  \displaystyle \left[\frac{d}{d\zeta} 
    \sum_{\xi_1 \in i\leftarrow b} \mWt_{\xi_1} 
    \sum_{\gamma\in\Gamma} \tilde{P}_{\gamma,i}'
    \sum_{\xi_2\in a\leftarrow\gamma} \mWt_{\xi_2} \right]_{\zeta=0}  \\
    \noalign{\medskip}
&=& \displaystyle \sum_{\xi_1 \in i\leftarrow b} \left[ \frac{d\mWt_{\xi_1}}{d\zeta} \right]_{\zeta=0}
    \sum_{\gamma\in\Gamma} P_{\gamma,i}'
    \sum_{\xi_2\in a\leftarrow\gamma} \mW_{\xi_2}  \\
    \noalign{\medskip}
&& \displaystyle +
    \sum_{\xi_1 \in i\leftarrow b} \mW_{\xi_1} 
    \sum_{\gamma\in\Gamma} \left[ \frac{d\tilde{P}_{\gamma,i}'}{d\zeta} \right]_{\zeta=0}\
    \sum_{\xi_2\in a\leftarrow\gamma} \mW_{\xi_2}   \\
    \noalign{\medskip}
&& \displaystyle +
    \sum_{\xi_1 \in i\leftarrow b} \mW_{\xi_1} 
    \sum_{\gamma\in\Gamma} P_{\gamma,i}' 
    \sum_{\xi_2\in a\leftarrow\gamma} \left[ \frac{d\mWt_{\xi_2}}{d\zeta} \right]_{\zeta=0},
    \end{array}
\end{equation}
where the tildes indicate that every branching probability $P_{\alpha,\beta}$ is
replaced by $P_{\alpha,\beta}e^{\xi\tau_\beta}$, as above.
The first and last terms are unchanged from graph $G$ in this construction, 
but the middle term,
\begin{equation}
	\begin{array}{lll}
&& \displaystyle \sum_{\xi_1 \in i\leftarrow b} \mW_{\xi_1} 
    \sum_{\gamma\in\Gamma} \left[ \frac{d\tilde{P}_{\gamma,i}'}{d\zeta} \right]_{\zeta=0}\
    \sum_{\xi_2\in a\leftarrow\gamma} \mW_{\xi_2}  \\
    \noalign{\medskip}
&=& \displaystyle \sum_{\xi_1 \in i\leftarrow b} \mW_{\xi_1}
  \sum_{\gamma\in\Gamma} \frac{P_{\gamma,x}P_{x,i}(\tau_i+\tau_x)+P_{\gamma,i}(\tau_i+P_{i,x}P_{x,i}\tau_x)}
                            {(1-P_{i,x}P_{x,i})^2}
  \sum_{\xi_2\in a\leftarrow\gamma} \mW_{\xi_2},
  	\end{array}
\end{equation}
is different (unless there is only one sink).
However, if we sum over sinks then $\sum_{a\in A}\sum_{\xi_2\in a\leftarrow\gamma} \mW_{\xi_2}=1$ for
all $\gamma$, and we can now simplify the sum over $\gamma$ as
\begin{equation}
	\sum_{\gamma\in\Gamma} \frac{P_{\gamma,x}P_{x,i}(\tau_i+\tau_x)+P_{\gamma,i}(\tau_i+P_{i,x}P_{x,i}\tau_x)}
						   {(1-P_{i,x}P_{x,i})^2} = \tau_i' = \sum_{\gamma\in\Gamma} P_{\gamma,i}'\tau_i'.
\end{equation}
The same argument can be applied whenever a trajectory reaches a node adjacent to $x$,
so that $\mT^G_b=\mT^{G'}_b$, as required.

The above procedure extends the approach of Bortz, Kalos and Lebowitz\cite{bortzkl75} 
to exclude not only the transitions from the current state into itself but also transitions
involving an adjacent node $x$.
Alternatively, this method could be viewed as a coarse-graining of the Markov chain.
Such coarse-graining is acceptable if the
property of interest is the average of the distribution of times rather than
the distribution of times itself. In our simulations the average is the only
property of interest. In cases when the distribution itself is sought, the approach
described here may still be useful and could be the first step
in the analysis of the distribution of escape times, as it yields the exact average
of the distribution.

\begin{figure}[ht]
\centerline{\includegraphics{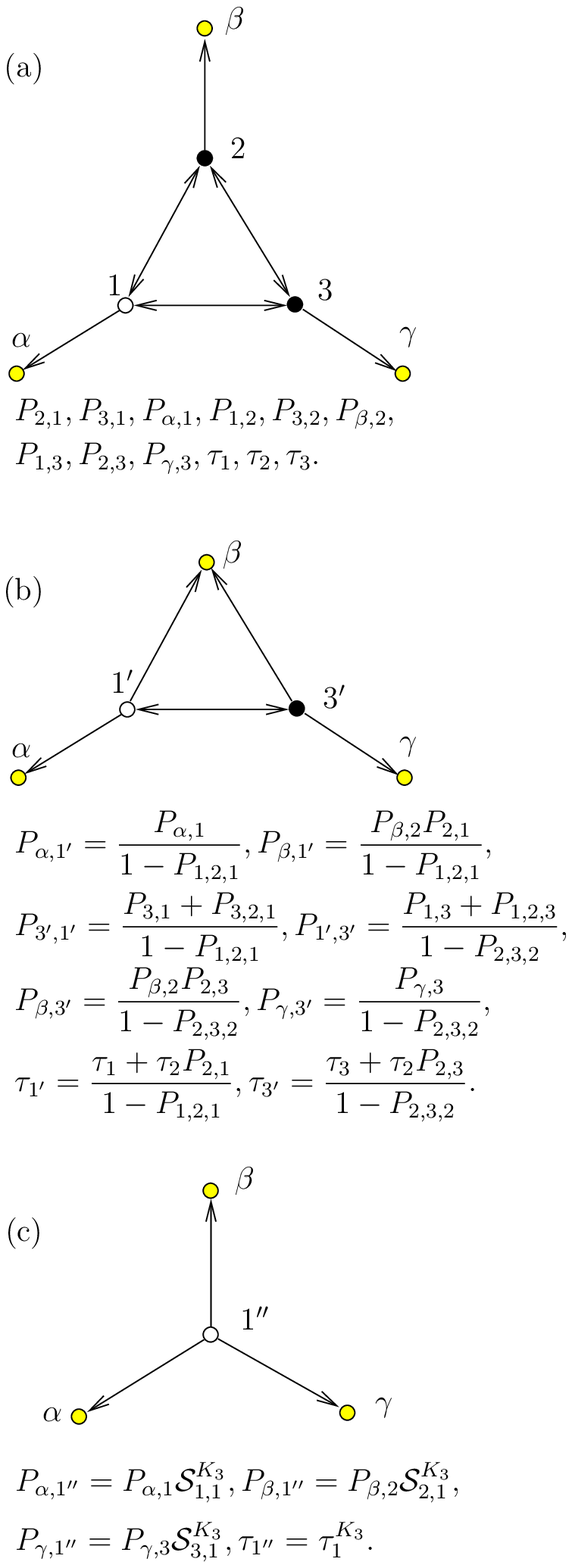}}
\caption{}
\label{fig:graph_transformation}
\end{figure}

The transformation is illustrated in Fig.~\ref{fig:graph_transformation}
for the case of a single source.
Fig.~\ref{fig:graph_transformation}~(a) displays the original graph and its parametrisation.
During the first iteration of the algorithm node $2$ is removed to yield the graph depicted in
Fig.~\ref{fig:graph_transformation}~(b).
This change 
reduces the dimensionality of the original graph, as the new graph contains one node and three edges 
fewer. The result of the second, and final, iteration of the algorithm is a graph that 
contains source and sink nodes only,
with the correct transition probabilities and mean waiting time 
(Fig.~\ref{fig:graph_transformation}~(c)).

We now describe algorithms to implement the above approach and calculate mean escape times from
complete graphs with multiple sources and sinks.
We follow the notation of \S\ref{sec:kmcanddps} and
consider a digraph $\mathcal{G}_N$ consisting of
$N_B$ source nodes, $N_A$ sink nodes, and $N_I$ intervening nodes.
$\mathcal{G}_N$ therefore contains the subgraph $G_{N_I+N_B}$.

The result of the transformation of a graph with a single source $b$ and $N_A$ sinks
using Algorithm~\ref{alg:CGT} is the mean escape time $\mT_b^{G_{N_I+1}}$
and $N_A$ pathway probabilities $\mathcal{P}_{\xi}$, $\xi\in A\leftarrow b$.
Solving a problem with $N_B$ sources is equivalent to solving $N_B$ single source problems.
For example, if there are two sources $b_1$ and $b_2$ we first solve a problem where only node
$b_1$ is set to be the source to obtain $\mT_{b_1}^{G_{N_I+N_B}}$
and the pathway sums from $b_1$ to every sink node $a \in A$.
The same procedure is then followed for $b_2$.

The form of the transition probability matrix ${\bf P}$ is illustrated below
at three stages: first for the original graph,
then at the stage when all the intervening nodes have been removed
(line \ref{line:DetachNodesf} in Algorithm~\ref{alg:CGT}),
and finally at the end of the procedure:
\begin{equation}
	\left(
	\begin{array}{c|c|c}
               {\bf 0} & A \leftarrow I	  & A \leftarrow B \\
	\hline {\bf 0} & I \leftrightarrows I & I \leftarrow B \\
	\hline {\bf 0} & B \leftarrow I	  & B \leftrightarrows B \\
	\end{array}
	\right)
	\rightarrow
	\left(
	\begin{array}{c|c|c}
		  {\bf 0} & {\bf 0} & A \leftarrow B \\
	\hline {\bf 0} & {\bf 0} & {\bf 0} \\
	\hline {\bf 0} & {\bf 0} & B \leftrightarrows B \\
	\end{array}
	\right)
	\rightarrow
	\left(
	\begin{array}{c|c|c}
		  {\bf 0} & {\bf 0} & A \leftarrow B \\
	\hline {\bf 0} & {\bf 0} & {\bf 0} \\
	\hline {\bf 0} & {\bf 0} & {\bf 0} \\
	\end{array}
	\right),
\end{equation}
Each matrix is split into blocks that specify the
transitions between the nodes of a particular type, as labelled. Upon termination,
every element in the top right block of matrix ${\bf P}$ is non-zero.

Algorithm~\ref{alg:CGT} uses the adjacency
matrix representation of graph $\mathcal{G}_N$, for which the average of the distribution
of mean first-passage times is to be obtained. 
For efficiency, when constructing the transition probability matrix ${\bf P}$
we order the nodes with the sink nodes first and the source nodes last. 
Algorithm~\ref{alg:CGT}
is composed of two parts. The first part (lines \ref{line:DetachNodess}-\ref{line:DetachNodesf})
iteratively removes all the intermediate
nodes from graph $\mathcal{G}_N$ to yield a graph that is composed of sink nodes and source nodes only.
The second part (lines \ref{line:Disconnects}-\ref{line:Disconnectf})
disconnects the source nodes from each other to produce
a graph with $N_A+N_B$ nodes and $(N_A+N_B)^2$ directed edges connecting each source
with every sink.
Finally, we evaluate the mean first-passage time for each source using the transformed graph.

The computational complexity of lines \ref{line:DetachNodess}-\ref{line:DetachNodesf}
of Algorithm~\ref{alg:CGT} is $\mathcal{O}(N_I^3 + N_I^2 N_B + N_I^2 N_A + N_I N_B^2 + N_I N_B N_A)$.
The second part of Algorithm~\ref{alg:CGT} (lines \ref{line:Disconnects}-\ref{line:Disconnectf}) scales
as $\mathcal{O}(N_B^3 + N_B^2 N_A)$.
The total complexity for the case of a single source
and for the case when there are no intermediate nodes
is $\mathcal{O}(N_I^3 + N_I^2 N_A)$ and $\mathcal{O}(N_B^3 + N_B^2 N_A)$, respectively.
The storage requirements of Algorithm~\ref{alg:CGT} scale as $\mathcal{O}\left((N_I+N_B)^2\right)$.

\begin{figure}[ht]
\centerline{\includegraphics{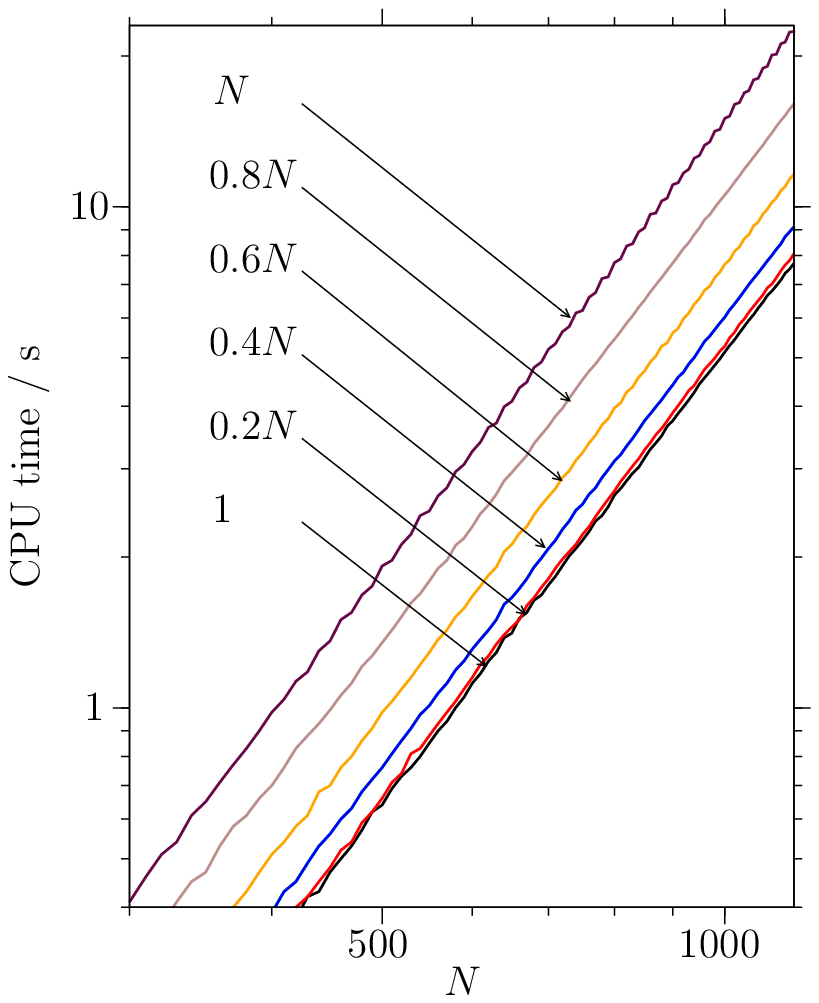}}
\caption{}
\label{fig:KnCPUofN}
\end{figure}

We have implemented the algorithm in Fortran 95 and timed it for complete graphs
of different sizes. The results presented in Fig.~\ref{fig:KnCPUofN} confirm the overall cubic scaling.
The program is GPL-licensed \cite{gpl} and available online.\cite{trygubcomgt}
These and other benchmarks presented in this work were obtained
for a single Intel$^{\text\textregistered}$ Pentium$^{\text\textregistered}$~4 3.00\,GHz
512\,Kb cache processor running under the Debian GNU/Linux operating system.\cite{debian}
The code was compiled and optimised using the Intel$^{\text\textregistered}$ Fortran compiler for Linux.

Finally, we note that the GT method described above
is a more general version of the method we introduced in our previous publication~\cite{TrygubenkoW04}.
It can be used to treat problems with multiple sources, which may be interconnected, thereby
allowing us to calculate KMC rate constants exactly. Furthermore, as described in \S\ref{sec:STO}, it can easily be extended to
treat problems where sets of sources and sinks overlap, allowing us to calculate the SS rate
constants in equation (\ref{eq:kABDPS}).

\section{Applications to sparse random graphs}
\label{sec:RN}

Algorithm~\ref{alg:CGT} could easily be adapted
to use adjacency-lists-based data structures \cite{CormenLRS01},
resulting in a faster execution and lower storage requirements for sparse graphs.
We have implemented \cite{trygubcomgt} a sparse-optimised version of Algorithm~\ref{alg:CGT}
because the graph representations of the Markov chains
of interest in the present work are sparse~\cite{chartrand77}.

The algorithm for detaching a single intervening node from an arbitrary graph
stored in a sparse-optimised format is given in Algorithm~\ref{alg:DetachNode}.
Having chosen the node to be removed, $\gamma$,
all the neighbours $\beta \in Adj[\gamma]$ are analysed in turn, as follows.
Lines \ref{line:findcs}-\ref{line:findcf} of Algorithm~\ref{alg:DetachNode}
find node $\gamma$ in the adjacency list of node $\beta$.
If $\beta$ is not a sink, lines \ref{line:notsinks}-\ref{line:notsinkf}
are executed to modify the adjacency list of node $\beta$: 
lines \ref{line:dels}-\ref{line:delf} delete node $\gamma$ from the
adjacency list of $\beta$, while lines \ref{line:adds}-\ref{line:addf}
make all the neighbours $\alpha \in Adj[\gamma]\ominus{\beta}$ of node $\gamma$ the neighbours of $\beta$.
The symbol $\ominus$ denotes the union minus the intersection of two sets, otherwise
known as the symmetric difference.
If the edge $\beta \rightarrow \alpha$ already existed
only the branching probability is changed (line \ref{line:pupdate}).
Otherwise, a new edge is created and the adjacency and branching
probability lists are modified accordingly
(lines \ref{line:news} and \ref{line:newf}, respectively).
Finally, the branching probabilities of node $\beta$ are
renormalised (lines \ref{line:renorms}-\ref{line:renormf}) and the waiting time for node $\beta$
is increased (line \ref{line:notsinkf}).

Algorithm~\ref{alg:DetachNode} is invoked iteratively for every node that is neither a
source nor a sink
to yield a graph that is composed of source nodes and sink nodes only. Then the procedure
described in \S\ref{sec:tauKN} for disconnection of source nodes
(lines \ref{line:Disconnects}-\ref{line:Disconnectf} of Algorithm~\ref{alg:CGT})
is applied to obtain the mean escape times for every source node. The sparse-optimised version
of the second part of Algorithm~\ref{alg:CGT} is straightforward 
and is therefore omitted here for brevity.

The running time of Algorithm~\ref{alg:DetachNode} is $\mathcal{O}(d_c \sum_{i \in Adj[c]} d_i)$,
where $d_k$ is the degree of node $k$. For the case when all the nodes in a graph have
approximately the same degree, $d$, the complexity is $\mathcal{O}(d^3)$.
Therefore, if there are $N$ intermediate nodes to be detached, and $d$ is of the same order of
magnitude as $N$, the cost of detaching $N$ nodes is $\mathcal{O}(N^4)$.
The asymptotic bound is worse than that of Algorithm~\ref{alg:CGT} because of the
searches through adjacency lists (lines \ref{line:findcs}-\ref{line:findcf}
and lines \ref{line:findas}-\ref{line:findaf}). If $d$ is sufficiently small the algorithm
based on adjacency lists is faster.

After each invocation of Algorithm~\ref{alg:DetachNode}
the number of nodes is always decreased by one.
The number of edges, however, can increase or decrease
depending on the in- and out-degree of the node to be removed
and the connectivity of its neighbours.
If node $\gamma$ is not directly connected to
any of the sinks, and the neighbours of node $\gamma$
are not connected to each other directly,
the total number of edges is increased by $d_\gamma(3-d_\gamma)$.
Therefore, the number of edges decreases (by $2$) only when
$d_\gamma \in \{1,2\}$, and the number of edges does not change if the degree is $3$.
For $d_\gamma>3$ the number of edges increases by an amount that grows quadratically with $d_\gamma$.
The actual increase depends on how many connections
already existed between the neighbours of $\gamma$.

The order in which the intermediate nodes are detached does not change the final result
and is unimportant if the graph is complete.
For sparse graphs, however, the order can affect the running time significantly.
If the degree distribution for successive graphs is sharp with the same average, $d$,
then the order in which the nodes are removed does not affect the complexity, which
is $\mathcal{O}(d^3 N)$.
If the distributions are broad it is helpful to remove the nodes with
smaller degrees first. A
Fibonacci heap min-priority queue \cite{fredmant87,CormenLRS01} was successfully
used to achieve this result. The overhead for maintaining a heap is $d_\gamma$ increase-key
operations (of $\mathcal{O}(\log (N))$ each) per execution
of Algorithm~\ref{alg:DetachNode}.
Fortran and Python implementations of Algorithm~\ref{alg:DetachNode}
algorithm are available online.\cite{trygubcomgt}

Random graphs provide an ideal testbed for the GT algorithm by providing control over the graph density.
A random graph, $R_N$, is obtained by starting with a set of $N$ nodes
and adding edges between them at random.\cite{wikipedia}
In this work we used a random graph model where
each edge is chosen independently with probability $\left<d\right>/(N-1)$,
where $\left<d\right>$ is the target value for the average degree.

\begin{figure}[ht]
\centerline{\includegraphics[width=0.7\textwidth]{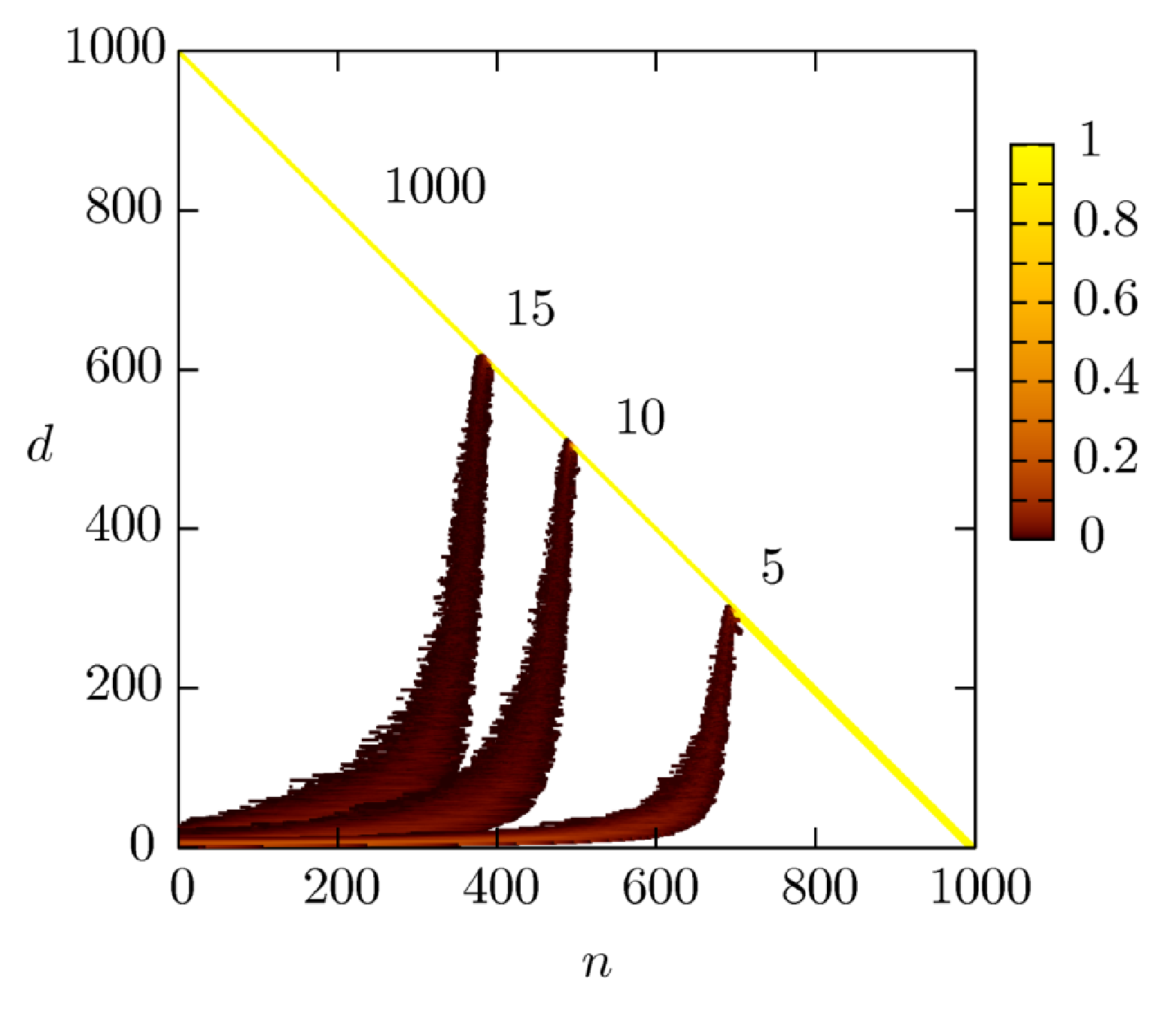}}
\caption{}
\label{fig:devolution}
\end{figure}

The complexity for removal of $N$ nodes can then be expressed as
\begin{equation}
\mathcal{O}\left(\log(N)\sum_{i \in \{1,2,3,\dots,N\}} 
\left( d_{c(i)}^2 \sum_{j \in Adj[c(i)]} d_{j,c(i)}\right)\right),
\label{eq:cost}
\end{equation}
where $d_{c(i)}$ is the degree of the node, $c(i)$, removed at iteration $i$, $Adj[c(i)]$ is its 
adjacency list, and $d_{j,c(i)}$ is the degree of the $j$th neighbour of that node at iteration $i$.
The computational cost given in Eq.~(\ref{eq:cost}) is difficult to express
in terms of the parameters of the original graph, as the cost of every cycle depends
on the distribution of degrees, the evolution of which, in turn, is dependent
on the connectivity of the original graph in a non-trivial manner (see Fig.~\ref{fig:devolution}).
The storage requirements of a sparse-optimised version of GT algorithm scale linearly with
the number of edges.

\begin{figure}[ht]
\centerline{\includegraphics{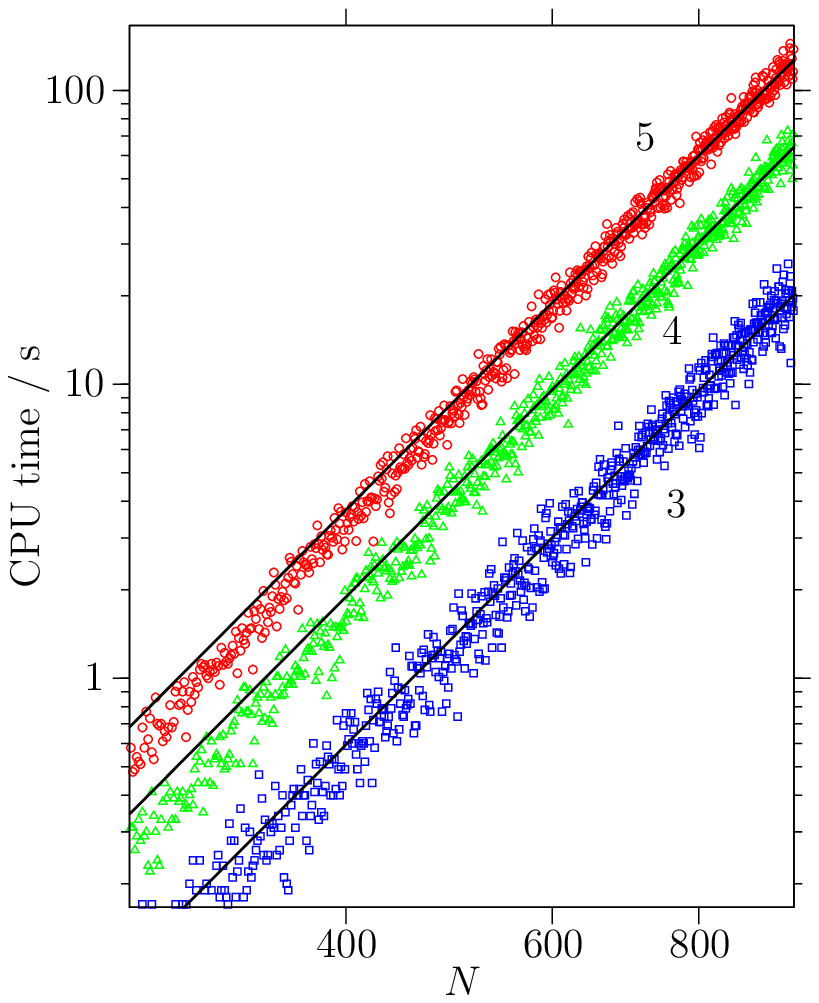}}
\caption{}
\label{fig:CPUofN}
\end{figure}

To investigate the dependence of the cost of the GT method on the number of nodes, $N$,
we have tested it on a series of random graphs $R_N$ for different values of $N$ and 
fixed average degree, $\left<d\right>$.
The results for three different values of $\left<d\right>$ are shown in Fig.~\ref{fig:CPUofN}.
The motivation for choosing $\left<d\right>$ from the interval $\left[3,5\right]$ was the fact
that most of our stationary point databases have average connectivities for the local minima
that fall into this range.

\begin{figure}[ht]
\centerline{\includegraphics{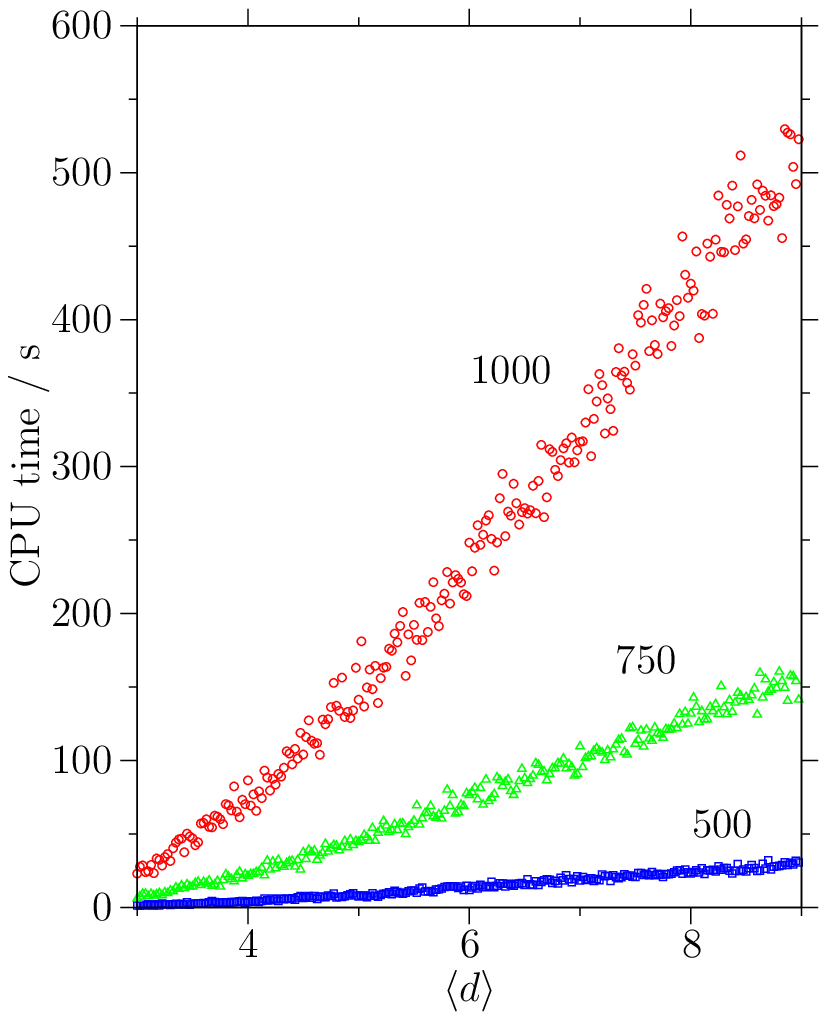}}
\caption{}
\label{fig:CPUofd}
\end{figure}

It can be seen from Fig.~\ref{fig:CPUofN}
that for sparse random graphs $R_N$ the cost scales as $\mathcal{O}(N^4)$ with a small
$\left<d\right>$-dependent prefactor.
The dependence of the computational complexity on $\left<d\right>$ is illustrated in Fig.~\ref{fig:CPUofd}.

\begin{figure}[ht]
\centerline{\includegraphics{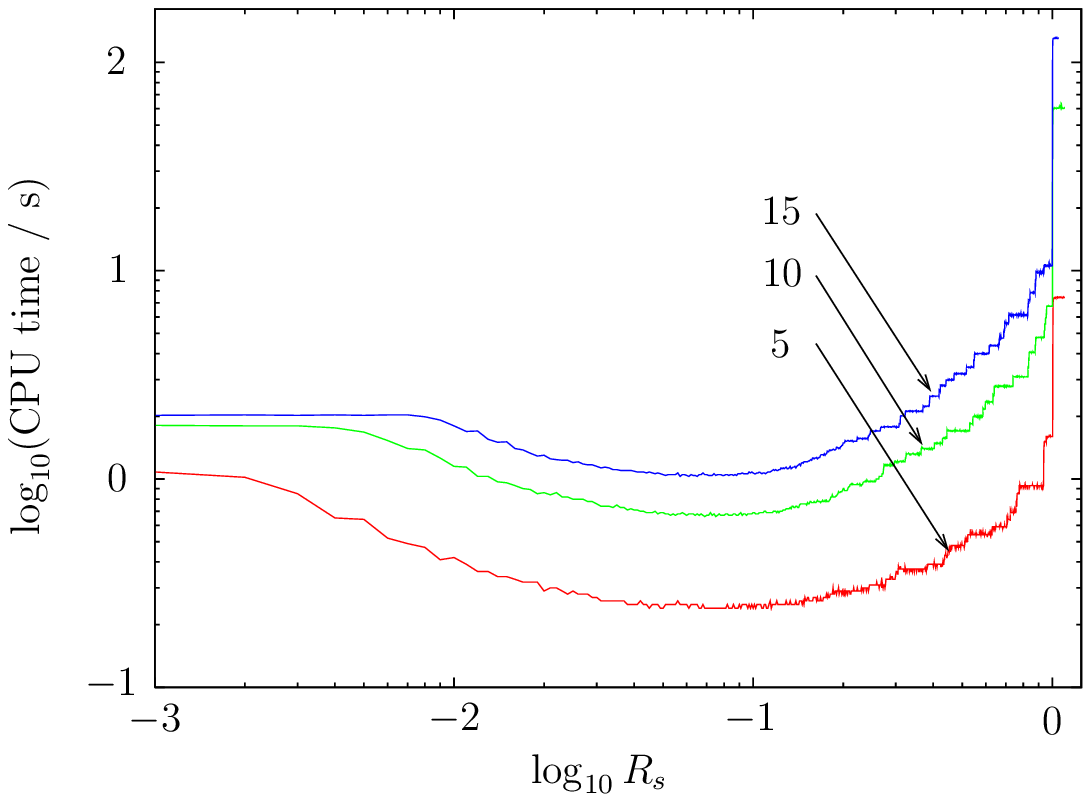}}
\caption{}
\label{fig:CPUofRs}
\end{figure}

From Fig.~\ref{fig:devolution} it is apparent that at some point
during the execution of the GT algorithm the graph reaches its maximum possible density.
Once the graph is close to complete
it is no longer efficient to employ a sparse-optimised algorithm.
The most efficient approach we have found for
sparse graphs is to use the sparse-optimised GT algorithm (SGT)
until the graph is dense enough, and then switch to dense-optimised GT
method (DGT), for which pseudocode is given in Algorithm~\ref{alg:CGT}.
We will refer to this approach as SDGT.
The change of data structures 
constitutes a negligible fraction of the total execution time.
Fig.~\ref{fig:CPUofRs} depicts the dependence of the CPU time as a function of the
switching parameter $R_s$.
Whenever the ratio $d_{c(i)}/n(i)$, where 
$n(i)$ is the number of nodes on a heap at iteration $i$,
is greater than $R_s$, the partially transformed graph is converted from the adjacency list format into
adjacency matrix format, and the transformation is continued using Algorithm~\ref{alg:CGT}.
It can be seen from Fig.~\ref{fig:devolution} that for the
case of a random graphs with a single sink, a single source and 999 intermediate nodes
the optimal values of $R_s$ lie in the interval $[0.07,0.1]$.

\section{Overlapping Sets of Sources and Sinks}
\label{sec:STO}

We now return to the digraph representation of a Markov chain that corresponds to the steady-state results
discussed in \S\ref{sec:kmcanddps}. A problem with overlapping sets of sources and sinks
can easily be converted into an equivalent problem where there is no overlap, and then the GT method
discussed in \S\ref{sec:tauKN} and \S\ref{sec:RN} can be applied as normal. Hence we can
evaluate the SS rate constants in equation (\ref{eq:kABDPS}) in a deterministic manner.

As discussed above, solving a problem with $n$ sources
reduces to solving $n$ single-source problems.
We can therefore explain how to deal with
a problem of overlapping sets of sinks and sources
for a simple example where there is a single source-sink $i$
and, optionally, a number of sink nodes.

First, a new node $i'$ is added to the
set of sinks and its adjacency lists are initialised to
$AdjOut[i']=\emptyset$ and $AdjIn[i']=AdjIn[i]$. Then, for every node $j\in AdjOut[i]$
we update its waiting time as $\tau_j=\tau_j+\tau_i$ and add node $j$ to the set of sources
with probabilistic weight initialised to $P_{j,i}W_i$,
where $W_i$ is the original probabilistic weight of source $i$
(the probability of choosing source $i$ from the set of sources).
Afterwards, the node $i$ is deleted.

\section{Applications to Lennard-Jones Clusters}
\label{sec:LJ}

\subsection{$O_h\leftrightarrow I_h$ Isomerisation of LJ$_{38}$}
\label{sec:LJ38}
We have applied the GT method to study the temperature dependence of the rate
of interconversion between configurations based on truncated octahedra ($O_h$)
and icosahedral ($I_h$) packing in a
38-atom Lennard-Jones cluster (LJ$_{38}$). The PES sample
was taken from a previous study~\cite{Wales02}
and contained 1740 minima and 2072 transition states, not including permutation-inversion isomers.
The assignment was made in Ref.~\citen{Wales02} by solving a master equation numerically
to find the eigenvector that corresponds to the smallest magnitude non-zero eigenvalue.
As simple two-state dynamics are associated with exponential rise and decay of occupation probabilities
there must exist a time scale on which all the exponential contributions to the solution
of the master equation decay to zero except for the slowest~\cite{Wales03}.
The sign of the components of the eigenvector corresponding to the slowest mode
was used to classify the minima as $I_h$ or $O_h$ in character~\cite{Wales02}.

The above sample was pruned to
remove disconnected minima and
create a digraph representation
that contained 759 nodes with 43 sources, 5 sinks, and 2639 edges.
The minimal, average and maximal degree for this graph were 2, $3.8$ and 84, respectively,
and the graph density was \mbox{$4.6\times10^{-3}$}.
We have used the SDGT algorithm with the switching ratio set to $0.08$ to transform this graph for several
values of temperature. In each of these calculations
622 out of 711 intermediate nodes were detached using SGT, and the remaining 89 intermediate nodes were detached using
the GT algorithm optimised for dense graphs (DGT).

\begin{figure}[ht]
\centerline{\includegraphics{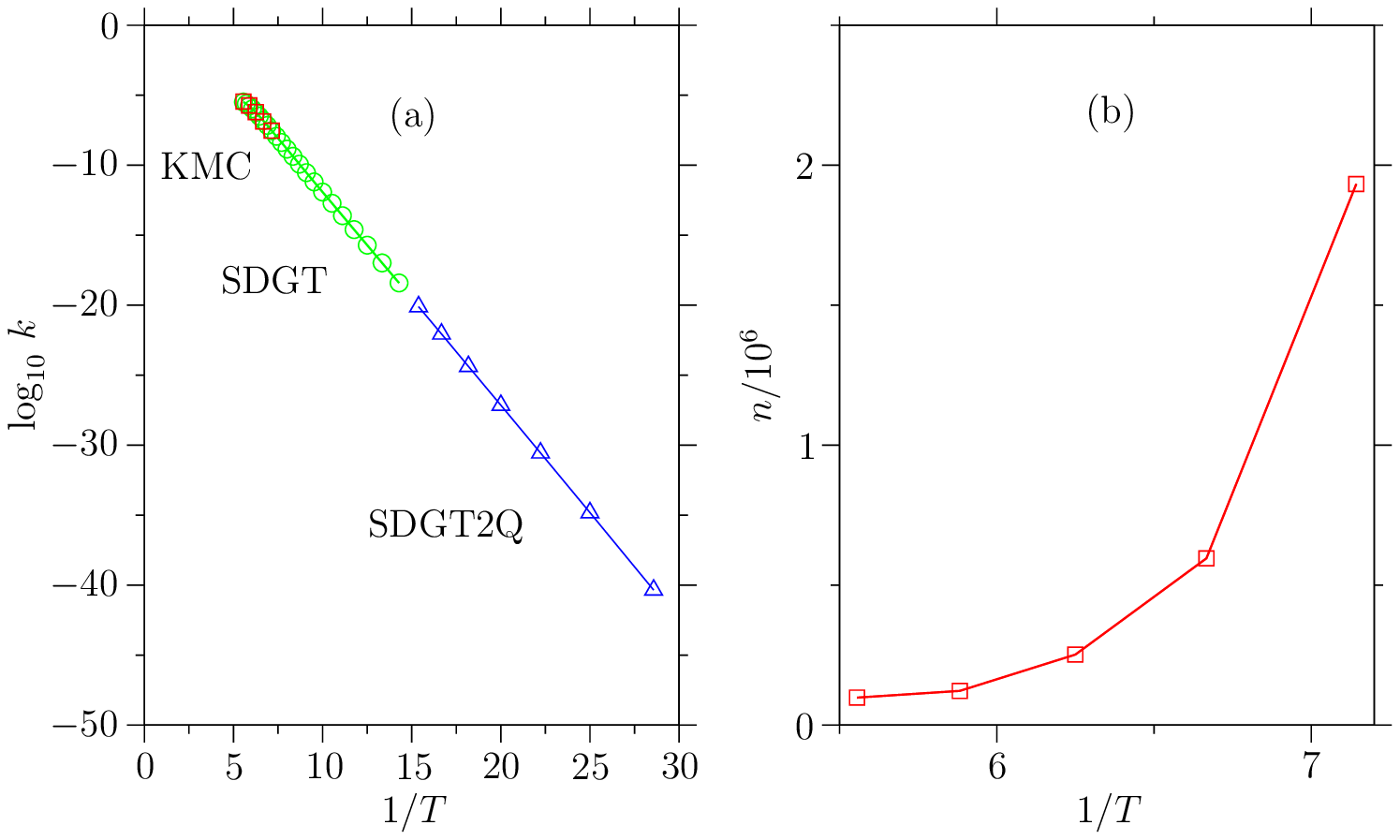}}
\caption{}
\label{fig:arrhenius38}
\end{figure}

An Arrhenius plot depicting the dependence of the rate constant on temperature
is shown in Fig.~\ref{fig:arrhenius38}~(a).
The running time for the SDGT algorithm was \mbox{$1.8\times10^{-2}$} seconds
[this value was obtained by averaging over 10 runs and was the same for each SDGT run in Fig.~\ref{fig:arrhenius38}~(a)].
For comparison, the timings obtained using the SGT and DGT algorithms for the same problem
were \mbox{$2.0\times10^{-2}$} and \mbox{$89.0\times10^{-2}$} seconds, respectively.
None of the 43 total escape probabilities (one for every source) deviated from unity
by more than $10^{-5}$ for temperatures above $T=0.07$ (reduced units).
For lower temperatures the probability was not conserved due to numerical imprecision.

The data obtained using SDGT method was compared with results from KMC simulation,
which require increasingly more CPU time as the temperature is lowered.
Fig.~\ref{fig:arrhenius38} also shows the data for KMC simulations at temperatures $0.14$, $0.15$, $0.16$,
$0.17$ and $0.18$. Each KMC simulation consisted of 1000 separate KMC trajectories
from which the averages were computed. The cost of each KMC calculation is proportional to the average
trajectory length, which is depicted in Fig.~\ref{fig:arrhenius38}~(b) as a function of the inverse temperature.
The CPU timings for each of these calculations were (in order of increasing temperature, averaged over five
randomly seeded KMC runs of 1000 trajectories each): $125$, $40$, $18$, $12$, and $7$ seconds.
It can be seen that using the GT method we were able to obtain kinetic data for a wider range of temperatures and
with less computational expense.

For the temperatue range of $[0.09,0.18]$ Arrhenius plots for this system were
obtained previously using KMC and master equation approaches, as well as the MM method~\cite{Wales02,Wales04}.
The results reported here are in quantitative agreement with these from Ref.~\citen{Wales04}.

\subsection{Internal Diffusion in LJ$_{55}$}
\label{sec:LJ55}

We have applied the graph transformation method to study the centre-to-surface atom migration in 55-atom
Lennard-Jones cluster. The global potential energy minimum for LJ$_{55}$ is
a Mackay icosahedron, which exhibits special stability and `magic number' properties~\cite{LabastieW90,BerryBDJ88}.
Centre-to-surface and surface-to-centre rates of migration of a tagged atom for this system were considered
in previous work~\cite{Wales04,TrygubenkoW06}. In Ref.~\citen{Wales04} the standard DPS procedure was applied to create
a stationary point database based on paths linking the global minimum with the tagged atom occupying the central position
to the same structure with the tagged atom in a surface site
(There are two inequivalent surface sites lying on two-fold and five-fold
axes of rotation).
We have reused this database in the present work.

The sample contained 9907 minima and 19384 transition states.
We excluded transition states corresponding to degenerate rearrangements
from consideration because they do not affect the rates~\cite{Wales04}.
For minima interconnected by more than one transition state
we added the rate constants to calculate the branching probabilities.
Four digraph representations were created with minimum degrees of 1, 2, 3 and 4
via iterative removal of the nodes with degrees that did not satisfy the requirement.
These digraphs will be referred to as digraphs 1, 2, 3 and 4, respectively.
The corresponding parameters are summarised in Table~\ref{tab:params}.
We quote
CPU timings for the DGT, SGT and SDGT methods for each of these graphs
in the last three columns of Table~\ref{tab:params}.
Each digraph contained two source nodes labelled $1$ and $2$ and a single sink.
The sink node corresponds to the global minimum with the tagged atom in the centre.
It is noteworthy that the densities of the graphs corresponding
to both our examples (LJ$_{38}$ and LJ$_{55}$) are significantly lower
than the values predicted for a complete sample~\cite{DoyeM05}, which makes
the use of sparse-optimised methods even more advantageous.
From Table~\ref{tab:params} it is clear that the SDGT approach is the fastest, as expected;
we will use SDGT for all the rate calculations in the rest of this section.

For this example KMC calculations are unfeasible at temperatures lower than about $T=0.3$
(reduced units throughout). Already for
$T=0.4$ the average KMC trajectory length is \mbox{$7.5\times 10^{6}$} (value obtained by averaging
over 100 trajectories). In Ref.~\citen{Wales04} it was therefore necessary to use the
steady-state approximation for the intervening minima to calculate the rate constant
at temperatures below $0.35$.
These results were not fully converged, and were revised in Ref.~\citen{TrygubenkoW06}
using alternative formulations of the kinetics.
Here we report results that are in direct correspondence with the KMC rate constants,
for temperatures as low as $0.1$.

\begin{figure}[ht]
\centerline{\includegraphics{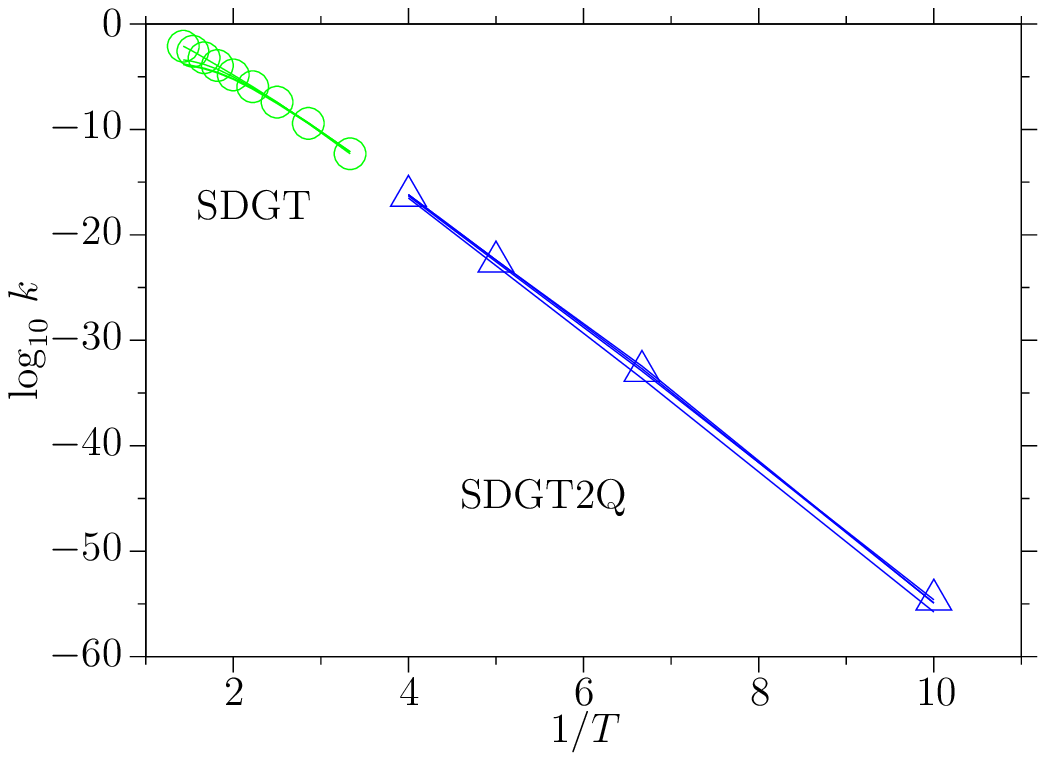}}
\caption{}
\label{fig:arrhenius55}
\end{figure}

Fig.~\ref{fig:arrhenius55} presents Arrhenius plots for rate constants calculated using the SDGT method.
The data denoted with circles are the results from seven SDGT calculations
at temperatures $T\in\{0.3,0.35,\dots,0.6\}$ conducted for each of the digraphs.
The total escape probabilities, $\Sigma_{1}^G$ and $\Sigma_{2}^G$, calculated for each of the two sources
at the end of the calculation deviated
from unity by no more than $10^{-5}$. For higher temperatures and
smaller digraphs the deviation was lower, being on the order of $10^{-10}$ for
digraph 4 at $T=0.4$.

At temperatures lower than $0.3$ the probability deviated by more than $10^{-5}$ due to numerical imprecision.
This problem was partially caused by the round-off errors in evaluation of the terms $1-P_{\alpha,\beta}P_{\beta,\alpha}$, which
increase when $P_{\alpha,\beta}P_{\beta,\alpha}$ approaches unity.
These errors can propagate and amplify as the evaluation proceeds. By writing
\begin{equation}
	\begin{array}{rll}
	P_{\alpha,\beta} &=& \displaystyle 1-\sum_{\gamma\not=\alpha}P_{\gamma,\beta}\equiv 1-\epsilon_{\alpha,\beta} \\
	\noalign{\medskip} {\rm and} \quad
	P_{\beta,\alpha} &=& \displaystyle 1-\sum_{\gamma\not=\beta}P_{\gamma,\alpha}\equiv 1-\epsilon_{\beta,\alpha},
	\end{array}
\end{equation}
and then using
\begin{equation}
	1-P_{\alpha,\beta}P_{\beta,\alpha}=\epsilon_{\alpha,\beta}-\epsilon_{\alpha,\beta}\epsilon_{\beta,\alpha}+\epsilon_{\beta,\alpha}
\end{equation}
we were able to decrease $1-\Sigma_{\alpha}^G$ by several orders of magnitude
at the expense of doubling the computational cost. The SDGT method with probability
denominators evaluated in this fashion will be referred to as SDGT1.

Terms of the form $1-P_{\alpha,\beta}P_{\beta,\alpha}$ approach zero
when nodes $\alpha$ and $\beta$ become `effectively' (i.e.~within available precision)
disconnected from the rest of the graph.
If this condition is encountered in the intermediate
stages of the calculation it could also mean that a larger subgraph of the original graph
is effectively disconnected.
The waiting time for escape if started from a node that belongs to this subgraph
tends to infinity. If the probability of getting to such a node from any of the source nodes is close to zero
the final answer may still fit into available precision, even though some of the intermediate values cannot.
Obtaining the final answer in such cases can be problematic as division-by-zero exceptions may occur.

Another way to alleviate numerical problems
at low temperatures is to stop round-off errors from propagation at early stages
by renormalising the branching probabilities of affected nodes $\beta\in Adj[\gamma]$
after node $\gamma$ is detached. The corresponding check that the updated probabilities of node $\beta$
add up to unity could be inserted after line~\ref{line:renormf} of Algorithm~\ref{alg:DetachNode} (see Appendix~\ref{app:algorithms}),
and similarly for Algorithm~\ref{alg:CGT}.
A version of SDGT method with this modification will be referred to as SDGT2.

Both SDGT1 and SDGT2 have similarly scaling overheads relative to the SDGT method.
We did not find any evidence for the superiority of one scheme over another.
For example, the SDGT calculation performed for digraph 4 at $T=0.2$
yielded $\mT^G=\mT^G_1 W_1 + \mT^G_2 W_2=6.4\times10^{-18}$,
and precision was lost as both $\Sigma_{1}^G$ and $\Sigma_{2}^G$ were less than
$10^{-5}$. The SDGT1 calculation resulted in $\mT^G=8.7\times10^{-22}$ and $\Sigma_{1}^G=\Sigma_{2}^G=1.0428$,
while the SDGT2 calculation produced $\mT^G=8.4\times10^{-22}$ with $\Sigma_{1}^G=\Sigma_{2}^G=0.99961$.
The CPU time required to transform this graph using our implementations of the SDGT1 and SDGT2 methods
was $0.76$ and $0.77$ seconds, respectively.

To calculate the rates at temperatures in the interval $[0.1,0.3]$ reliably
we used an implementation of the SDGT2 method
compiled with quadruple precision (SDGT2Q)
(note that the architecture is the same as
in other benchmarks, i.e.~with 32 bit wide registers).
The points denoted by triangles in Fig.~\ref{fig:arrhenius55} are the results from 4 SDGT2Q calculations
at temperatures $T\in\{0.10,0.35,\dots,0.75\}$. These results are in agreement 
with those previously reported for this system~\cite{TrygubenkoW04}.

Using the SDGT2Q formulation we were also able to
reach lower temperatures for the LJ$_{38}$ example
presented in the previous section.
The corresponding data is shown in Fig.~\ref{fig:arrhenius38}~(triangles).

\section{Conclusions}
The most important result of this work is probably the graph transformation (GT) method.
The method works with a digraph representation of a Markov chain
and can be used to calculate the first moment of a distribution of the first-passage times,
as well as the total transition probabilities for an arbitrary digraph with
sets of sources and sinks that can overlap.
The calculation is performed in a non-iterative and non-stochastic manner,
and the number of operations is independent of the simulation temperature.

We have presented three implementations of the GT algorithm: sparse-optimised (SGT),
dense-optimised (DGT), and hybrid (SDGT), which is a combination of the first two.
The SGT method uses a Fibonacci heap min-priority queue
to determine the order in which the intermediate
nodes are detached to achieve slower growth of the graph density and, consequently,
better performance.
SDGT is identical to DGT if the graph is dense. For sparse graphs SDGT performs
better then SGT because it switches to DGT when the density of a graph being transformed
approaches the maximum.
We find that for SDGT method performs well both for sparse and dense graphs.
The worst case asymptotic scaling of SDGT is cubic.

We have also suggested two versions of the SDGT method
that can be used in calculations where a greater degree of precision is required.
The code that implements SGT, DGT, SDGT, SDGT1 and SDGT2 methods
is available for download~\cite{trygubcomgt}.

The connection between the steady-state (SS) kinetic formulation
and KMC approaches was discussed in terms of digraph representations of Markov chains.
We showed that rate constants obtained using both the KMC or SS methods can be computed using graph
transformation. We have presented applications to the isomerisation of the LJ$_{38}$ cluster and internal
diffusion in the LJ$_{55}$ cluster.
Using the GT method we were able to calculate rate constants at lower temperatures
and with less computational expense.

We also obtained analytic expressions for the total transition probabilities in arbitrary digraphs
in terms of combinatorial sums over pathway ensembles. It is hoped that these results
will help in further work, for example, obtaining higher moments of the distribution
of the first-passage times for arbitrary Markov chains.

\section{Acknowledgements}
S.A.T.~is grateful to Cambridge Commonwealth Trust/Cambridge Overseas Trust
and Darwin College for the financial support.
Authors would like to thank Tim James, Dwaipayan Chakrabarti and Dr. David J. C. MacKay
for comments on the manuscript.

\bibliographystyle{thesis}
\bibliography{trygub}

\begin{thebibliography}{10}

\bibitem{bolchgmt98}
G.~Bolch, S.~Greiner, H.~de~Meer  and K.~S. Trivedi, \emph{Queueing Networks
  and {M}arkov Chains}, \wiley (1998).

\bibitem{GrimmettS05a}
G.~R. Grimmett and D.~R. Stirzaker, \emph{Probability and Random Processes},
  Oxford University Press, Oxford (2005).

\bibitem{GrimmettS05b}
G.~R. Grimmett and D.~R. Stirzaker, \emph{One Thousand Exercises in
  Probability}, Oxford University Press, Oxford (2005).

\bibitem{Wales03}
D.~J. Wales, \emph{Energy Landscapes: Applications to Clusters, Biomolecules
  and Glasses}, \cam (2003).

\bibitem{Wales02}
D.~J. Wales, \mp \textbf{100}, 3285 (2002).

\bibitem{kampen81}
N.~G. van Kampen, \emph{Stochastic Processes in Physics and Chemistry},
  \elsevier (1981).

\bibitem{bortzkl75}
A.~B. Bortz, M.~H. Kalos  and J.~L. Lebowitz, J. Comput. Phys. \textbf{17}, 10
  (1975).

\bibitem{fichthornw91}
K.~A. Fichthorn and W.~H. Weinberg, J. Chem. Phys \textbf{95}, 1090 (1991).

\bibitem{Miller99}
M.~A. Miller, \emph{Energy Landscapes and Dynamics of Model Clusters}, Ph.D.
  thesis, University of Cambridge (March 1999).

\bibitem{blockks04}
M.~Block, R.~Kunert, E.~Sch\"{o}ll, T.~Boeck  and T.~Teubner, \njp \textbf{6},
  166 (2004).

\bibitem{mukherjeesz03}
D.~Mukherjee, C.~G. Sonwane  and M.~R. Zachariah, \jcp \textbf{119}, 3391
  (2003).

\bibitem{bulnespr98}
F.~M. Bulnes, V.~D. Pereyra  and J.~L. Riccardo, \pre \textbf{58}, 86–92
  (1998).

\bibitem{kunz95}
R.~E. Kunz, \emph{Dynamics of First-Order Phase Transitions}, Deutsch, Thun
  (1995).

\bibitem{Wales04}
D.~J. Wales, \mp \textbf{102}, 883 (2004).

\bibitem{evansw04}
D.~A. Evans and D.~J. Wales, \jcp \textbf{121}, 1080 (2004).

\bibitem{evansw03}
D.~A. Evans and D.~J. Wales, \jcp \textbf{119}, 9947 (2003).

\bibitem{ApaydinBGHL02}
M.~S. Apaydin, D.~L. Brutlag, C.~Guestrin, D.~Hsu  and J.-C. Latombe, in
  \emph{RECOMB '02: Proceedings of the Sixth Annual International Conference on
  Computational Biology}, pp. 12--21. ACM Press, New York, NY (2002).

\bibitem{ApaydinGVBL02}
M.~S. Apaydin, C.~E. Guestrin, C.~Varma, D.~L. Brutlag  and J.-C. Latombe, \bi
  \textbf{18}, S18 (2002).

\bibitem{SinghalSP04}
N.~Singhal, C.~D. Snow  and V.~S. Pande, \jcp \textbf{121}, 415 (2004).

\bibitem{chartrand77}
G.~Chartrand, \emph{Introductory Graph Theory}, \dover (1977).

\bibitem{CormenLRS01}
T.~H. Cormen, C.~E. Leiserson, R.~L. Rivest  and C.~Stein, \emph{Introduction
  to Algorithms}, \mit, Cambridge, Mass., 2nd edn. (2001).

\bibitem{dads}
NIST, \emph{Dictionary of Algorithms and Data Structures},
  http://www.nist.gov/dads/ (2005).

\bibitem{wikipedia}
\emph{Wikipedia, the Free Encyclopedia}, http://www.wikipedia.org/
  (\currentyear).

\bibitem{Bar-haimK98}
A.~Bar-Haim and J.~Klafter, \jcp \textbf{109}, 5187 (1998).

\bibitem{novotny94}
M.~A. Novotny, \prl \textbf{74}, 1 (1995).

\bibitem{bulmer79}
M.~G. Bulmer, \emph{Principles of Statistics}, \dover (1979).

\bibitem{trumbo99}
B.~E. Trumbo, \emph{Relationship Between the {P}oisson and Exponential
  Distributions},
  http://www.sci.csuhayward.edu/statistics/Resources/Essays/PoisExp.htm (1999).

\bibitem{middleton03}
T.~F. Middleton, \emph{Energy Landscapes of Model Glasses}, Ph.D. thesis,
  University of Cambridge (March 2003).

\bibitem{reede81}
D.~A. Reed and G.~Ehrlich, \ss \textbf{105}, 603 (1981).

\bibitem{voter05}
A.~F. Voter, in \emph{Radiation Effects in Solids}, pp. 1--22. \springer
  (2005).

\bibitem{Raykin92}
M.~Raykin, \jpa \textbf{26}, 449 (1992).

\bibitem{MurthyK89}
K.~P.~N. Murthy and K.~W. Kehr, \pra \textbf{40}, 2082 (1989).

\bibitem{Trygubenko06}
S.~A. Trygubenko, \emph{Pathways and Energy Landscapes}, Ph.D. thesis,
  University of Cambridge (January 2006).

\bibitem{TrygubenkoW06}
S.~A. Trygubenko and D.~J. Wales, \mpns, in press  (2006).

\bibitem{GoldhirschG86}
I.~Goldhirsch and Y.~Gefen, \pra \textbf{33}, 2583 (1986).

\bibitem{GoldhirschG87}
I.~Goldhirsch and Y.~Gefen, \pra \textbf{35}, 1317 (1987).

\bibitem{weisstein04cg}
E.~W. Weisstein, \emph{`{C}omplete Graph'. {F}rom {M}athWorld --- {A} {W}olfram
  Web Resource}, http://mathworld.wolfram.com/CompleteGraph.html (2005).

\bibitem{goldberg60}
S.~Goldberg, \emph{Probability: An Introduction}, \dover (1960).

\bibitem{KahngR89}
B.~Kahng and S.~Redner, \jpa \textbf{22}, 887 (1989).

\bibitem{ZhengLW95}
D.~Zheng, Y.~Liu  and Z.~D. Wang, \jpa \textbf{28}, L409 (1995).

\bibitem{KimL95}
S.~K. Kim and H.~H. Lee, \jap \textbf{78}, 3809 (1995).

\bibitem{BressloffDK96}
P.~C. Bressloff, V.~M. Dwyer  and M.~J. Kearney, \jpa \textbf{29}, 1881 (1996).

\bibitem{RevathiBLM96}
S.~Revathi, V.~Balakrishnan, S.~Lakshmibala  and K.~P.~N. Murthy, \pre
  \textbf{54}, 2298 (1996).

\bibitem{KimCK98}
K.~Kim, J.~S. Choi  and Y.~S. Kong, \jpsj \textbf{67}, 1583 (1998).

\bibitem{KimKK00}
K.~Kim, G.~H. Kim  and Y.~S. Kong, Fractals \textbf{8}, 181 (2000).

\bibitem{AsikainenHA02}
J.~Asikainen, J.~Heinonen  and T.~Ala-Nissila, \pre \textbf{66}, 066706 (2002).

\bibitem{PuryC03}
P.~A. Pury and M.~O. C\'{a}ceres, \jpa \textbf{36}, 2695 (2003).

\bibitem{SlutskyKM04}
M.~Slutsky, M.~Kardar  and L.~A. Mirny, \pre \textbf{69}, 061903 (2004).

\bibitem{SlutskyM04}
M.~Slutsky and L.~A. Mirny, \bj \textbf{87}, 4021 (2004).

\bibitem{Ben-AvrahamRC89}
D.~ben Avraham, S.~Redner  and Z.~Cheng, \jsp \textbf{56}, 437 (1989).

\bibitem{GefenG89}
Y.~Gefen and I.~Goldhirsch, \Pd \textbf{38}, 119 (1989).

\bibitem{MatanH89}
O.~Matan and S.~Havlin, \pra \textbf{40}, 6573 (1989).

\bibitem{HauckeWBUW90}
H.~Haucke, S.~Washburn, A.~D. Benoit, C.~P. Umbach  and R.~A. Webb, \prb
  \textbf{41}, 12454 (1990).

\bibitem{RevathiB93a}
S.~Revathi and V.~Balakrishnan, \jpa \textbf{26}, 467 (1993).

\bibitem{NoskowiczG90}
S.~H. Noskowicz and I.~Goldhirsch, \pra \textbf{42}, 2047 (1990).

\bibitem{RevathiB93b}
S.~Revathi and V.~Balakrishnan, \pre \textbf{47}, 916 (1993).

\bibitem{BalakrishnanV95}
V.~Balakrishnan and C.~Vandenbroeck, \Pa \textbf{217}, 1 (1995).

\bibitem{KersteinP87}
A.~R. Kerstein and R.~B. Pandey, \prb \textbf{35}, 3575 (1987).

\bibitem{GefenG87}
Y.~Gefen and I.~Goldhirsch, \prb \textbf{35}, 8639 (1987).

\bibitem{GefenG85}
Y.~Gefen and I.~Goldhirsch, \jpa \textbf{18}, L1037 (1985).

\bibitem{HausK87}
J.~W. Haus and K.~W. Kehr, \prep \textbf{150}, 263 (1987).

\bibitem{NoskowiczG87}
I.~G. S.~H.~Noskowicz, \jsp \textbf{48}, 255 (1987).

\bibitem{LandauerB87}
R.~Landauer and M.~Buttiker, \prb \textbf{36}, 6255 (1987).

\bibitem{Tao87}
R.~Tao, \jpa \textbf{20}, 6151 (1987).

\bibitem{KoplikRW88}
J.~Koplik, S.~Redner  and D.~Wilkinson, \pra \textbf{37}, 2619 (1988).

\bibitem{dijk93}
N.~M. van Dijk, \emph{Queueing Networks and Product Forms}, \wiley (1993).

\bibitem{gelenbep98}
E.~Gelenbe and G.~Pujolle, \emph{Introduction to Queueing Networks}, \wiley
  (1998).

\bibitem{conwayg89}
A.~E. Conway and N.~D. Georganas, \emph{Queueing Networks - Exact Computational
  Algorithms}, \mit (1989).

\bibitem{eppsteingi97}
D.~Eppstein, Z.~Galil  and G.~F. Italiano, in \emph{Algorithms and Theory of
  Computation Handbook}, edited by M.~J. Atallah, chap.~8, CRC Press (1999).

\bibitem{cherkasskygr94}
B.~V. Cherkassky, A.~V. Goldberg  and T.~Radzik, in \emph{SODA '94: Proceedings
  of the Fifth Annual {ACM-SIAM} Symposium on Discrete Algorithms}, pp.
  516--525, Society for Industrial and Applied Mathematics, Philadelphia, PA
  (1994).

\bibitem{ramalingamr96a}
G.~Ramalingam and T.~Reps, J. Algorithms \textbf{21}, 267 (1996).

\bibitem{ramalingamr96b}
G.~Ramalingam and T.~Reps, Theor. Comput. Sci. \textbf{158}, 233 (1996).

\bibitem{gpl}
\emph{{GNU} General Public License}, http://www.gnu.org/copyleft/gpl.html
  (\currentyear).

\bibitem{trygubcomgt}
S.~A. Trygubenko, \emph{Graph Transformation Program},
  http://www.trygub.com/gt/ (\currentyear).

\bibitem{debian}
\emph{Debian --- The Universal Operating System}, http://www.debian.org/
  (\currentyear).

\bibitem{TrygubenkoW04}
S.~A. Trygubenko and D.~J. Wales, \jcp \textbf{120}, 2082 (2004).

\bibitem{fredmant87}
M.~L. Fredman and R.~E. Tarjan, Journal of the ACM \textbf{34}, 596 (1987).

\bibitem{LabastieW90}
P.~Labastie and R.~L. Whetten, \prl \textbf{65}, 1567 (1990).

\bibitem{BerryBDJ88}
R.~S. Berry, T.~L. Beck, H.~L. Davis  and J.~Jellinek, Adv. Chem. Phys.
  \textbf{70B}, 75 (1988).

\bibitem{DoyeM05}
J.~P.~K. Doye and C.~P. Massen, \jcp \textbf{122}, 084105 (2005).

\bibitem{pitman93}
J.~Pitman, \emph{Probability}, \springer (1993).

\end{thebibliography}
\clearpage

\section*{Figure Captions}
\begin{enumerate}
\item
BKL algorithm for propagating a KMC trajectory applied to a three-state Markov chain.
(a) The transition state diagram is shown where states and transitions are represented
by circles and directed arrows, respectively. The Markov chain is parametrised by
transition probabilities $P_{\alpha,1}$, $P_{\beta,1}$ and $P_{1,1}$.
Absorbing states $\alpha$ and $\beta$
are shaded. If $P_{1,1}$ is close to unity the KMC trajectory is likely
to revisit state $1$ many times before going to $\alpha$ or $\beta$.
(b) State $1$ is replaced with state $1'$. The new Markov chain is parametrised by
transition probabilities $P_{\alpha,1'}$, $P_{\beta,1'}$
and the average time for escape from state $1$ is $\tau_1$.
Transitions from state $1'$ to itself are forbidden. Every time state $1'$ is visited the
simulation `clock' is incremented by $\tau_1$.
\item
The computational cost of the kinetic Monte Carlo and matrix
multiplication methods as a function of escape probabilities for $K_3$.
$M$ is the number of matrix multiplications required to converge the value
of the total probability of getting from node $1$ to nodes $1$, $2$ and $3$:
the calculation was terminated when the change in the total probability between
iterations was less than $10^{-3}$. The number of matrix
multiplications $M$ and the average trajectory length $\left<l\right>$ can be used as a measure
of the computational cost of the matrix multiplication and kinetic Monte Carlo approaches,
respectively. The computational requirements of the exact summation method are
independent of $\mathcal{E}$. 
Note the log$_{10}$ scale on the vertical axis.
\item
The graph transformation algorithm of \S\ref{sec:tauKN} at work.
(a) A digraph with $6$ nodes and $9$ edges.
The source node is node $1$ (white), the sinks are nodes $\alpha$, $\beta$ and $\gamma$ (shaded),
and the intermediate nodes are $2$ and $3$ (black).
The waiting times and transition probabilities that parametrise the graph are given below
the diagram.
(b) Node $2$ and all its incoming and outgoing edges are deleted from the graph depicted in 
(a). Two edges $\beta \leftarrow 1$ and $\beta \leftarrow 3$ are added. 
The parameters for this new graph are denoted by primes and expressed in terms of the
parameters for the original graph.
(c) Node $3$ is now disconnected as well. 
The resulting graph is composed of source and sink nodes only. The total
probability and waiting times coincide with those of $K_3$, as expected. 
The new parameters are denoted by a double prime.
\item
CPU time needed to transform a dense graph $G_{2N}$ using Algorithm~\ref{alg:CGT}
as a function of $N$. The graph $G_{2N}$ is composed
of a $K_N$ subgraph and $N$ sink nodes. The data is shown for six different cases,
including a single source, and for sources comprising 20, 40, 60, 90,
and 100 percent of the nodes in $K_N$, as labelled.
The data for the cases $1$ and $N$ was fitted as $5.1\times10^{-9}N^3$ and $1.5\times10^{-8}N^3$,
respectively (curves not shown). For case $1$ only DetachNode operations were performed
while for $N$ only Disconnect operations were used.
Note the log$_{10}$ scale on both axes.
\item
Evolution of the distribution of degrees for random graphs of different expected degree, $\left<d\right>=5,10,15$, as
labelled. This is a colour-coded projection of the probability mass function,\cite{bolchgmt98,pitman93} $P(d)$, 
of the distribution of degrees as a function of the number of the detached intermediate nodes, $n$. The straight line
shows $P(d,n)$ for complete graph $K_{1000}$. All four graphs contain a single source, 999 intermediate nodes and
a single sink. The transformation was performed using a sparse-optimised version of Algorithm~\ref{alg:CGT}
with a Fibonacci-heap-based min-priority queue.
It can be seen that as the intermediate nodes are detached the density of the graph that is
being transformed grows. The expected degree of the initial graph determines how soon the maximum density will
be reached.
\item
CPU time needed to transform a sparse random graph $R_{2N}$ using the GT approach described in \S\ref{sec:tauKN}
as a function of the number of intermediate nodes, $N$. $R_{2N}$ is composed of a single source node, $N$ sink nodes
and $N-1$ intervening nodes. For each value of $N$ the data for three different values of the expected degree,
$\left<d\right>=3,4,5$, is shown, as labelled.
Solid lines are analytic fits of the form
$c N^4$, where $c=2.3\times10^{-11}, 7.4\times10^{-11}, 1.5\times10^{-10}$
for $\left<d\right>=3,4,5$, respectively. The CPU time is in seconds.
Note the log$_{10}$ scale on both axes.
\item
CPU time needed to transform a sparse random graph $R_{2N}$ using the GT approach
as a function of the expected degree, $\left<d\right>$. The data is shown
for three graphs with $N=500$, $750$ and $1000$, as labelled.
$R_{2N}$ is composed of a single source node, $N$ sink nodes
and $N-1$ intermediate nodes.
\item
CPU time as a function of switching ratio $R_s$ for random graphs of different expected degree,
$\left<d\right>=5,10,15$, as labelled.
All three graphs contain a single source, 999 intermediate nodes and
a single sink. The transformation was performed using the sparse-optimised version of Algorithm~\ref{alg:CGT}
until the the ratio $d_{c(i)}/n(i)$ became greater than $R_s$. Then a partially transformed graph
was converted into adjacency matrix format and transformation was continued with Algorithm~\ref{alg:CGT}.
The optimal value of $R_s$ lies in the interval $[0.07,0.1]$.
Note the log$_{10}$ scale on both axes.
\item
(a) Arrhenius plots for the LJ$_{38}$ cluster. $k$ is the rate constant corresponding to transitions
from icosahedral-like to octahedral-like regions. Green circles represent the data obtained from 23 SDGT runs
at temperatures $T\in\{0.07,0.075,\dots,0.18\}$. The data from five KMC runs is also shown (squares).
The data shown using triangles corresponds to temperatures $T\in\{0.035,0.04,\dots,0.065\}$ and
was obtained using the SDGT2 method (discussed in \S\ref{sec:LJ55}) with quadruple precision enabled.
In all SDGT runs the total escape probabilities calculated for every source at the end of the calculation
deviated from unity by no more then $10^{-5}$. For this stationary point database the lowest temperature for which
data was reported in previous works was $T=0.03$.
(b) The average KMC trajectory length [data in direct correspondence with KMC results shown in (a)]. A solid line is
used to connect the data points to guide the eye.
\item
Arrhenius plots for four digraphs of varying sizes
(see Table~\ref{tab:params}) created from a stationary point database for the LJ$_{55}$ cluster.
$k$ is the rate constant corresponding to
surface-to-centre migration of a tagged atom.
Calculations were conducted at $T\in\{0.3,0.35,\dots,0.7\}$ using the SDGT
method (circles) and $T\in\{0.1,0.15,\dots,0.25\}$ using SDGT2Q (triangles).
For each of the digraphs the calculations yielded essentially identical
results, so data points for only one of them are shown.
Lines connecting the data points are shown for each of the digraphs.
\end{enumerate}
\clearpage

\section*{Figures}

\clearpage

\clearpage

\clearpage

\clearpage

\clearpage

\clearpage

\clearpage

\clearpage

\clearpage

\clearpage
\section*{Tables}
\begin{table}[hb]
\caption{Properties of four digraphs corresponding to the LJ$_{55}$ PES sample from an internal diffusion study.
$|V|$ is the number of nodes; $|E|$ is the number of directed edges; $d_{min}$, $\left<d\right>$ and $d_{max}$
are the minimum, average and maximum degrees, respectively; $\rho$ is the graph density, defined as a ratio
of the number of edges to the maximum possible number of edges; $r$ and $d$ are the graph radius
and diameter, defined as the maximum and minimum node eccentricity, respectively,
where the eccentricity of a node $v$ is defined as the maximum distance between $v$ and any other node.
$\left<l\right>$ is the average distance between nodes.
The CPU time, $t$, necessary to transform each graph using
the DGT, SGT and SDGT methods is given in seconds for a single 32-bit
Intel$^{\text\textregistered}$ Pentium$^{\text\textregistered}$~4 3.00\,GHz 512\,Kb cache processor.
}
\label{tab:params}
\begin{center}
\begin{tabular}{cccr@{.}lcr@{.}lrr@{.}lrr@{.}lr@{.}lr@{.}l}
\hline
\hline
$|V|$&$|E|$&$d_{min}$&\multicolumn{2}{c}{$\left<d\right>$}&$d_{max}$&\multicolumn{2}{c}{$\rho/10^{-4}$}&$r$%
&\multicolumn{2}{c}{$\left<l\right>$}&$d$&\multicolumn{2}{c}{$t_{\rm DGT}$}%
&\multicolumn{2}{c}{$t_{\rm SGT}$}&\multicolumn{2}{c}{$t_{\rm SDGT}$}\\ 
\hline                                                                                                     
\multicolumn{1}{r}{9843}&\multicolumn{1}{r}{34871}&1& 3&9& 983&  3&6&10 &5&71 &20& 2346&1& 39&6& 1&36 \\ 
\multicolumn{1}{r}{6603}&\multicolumn{1}{r}{28392}&2& 4&8& 983&  6&5& 9 &4&86 &17& 1016&1& 38&9& 1&33 \\ 
\multicolumn{1}{r}{2192}&\multicolumn{1}{r}{14172}&3& 7&9& 873& 29&5& 4 &3&63 & 8&   46&9&  5&9& 0&49 \\ 
\multicolumn{1}{r}{ 865}&\multicolumn{1}{r}{ 7552}&4& 1&9& 680&101&0& 4 &3&07 & 7&    3&1&  0&8& 0&12 \\ 
\hline
\hline
\end{tabular}
\end{center}
\end{table}

\clearpage
\section*{Algorithms}
\label{app:algorithms}
\renewcommand\baselinestretch{1}

\begin{algorithm}
\caption{Calculate the pathway sum $\mathcal{S}_{a,b}^{G_N}$ for an arbitrary digraph $G_N$}
\label{alg:ProbForKn}
\begin{algorithmic}[1]
\REQUIRE{$1 < N$ and $a,b \in \{0,1,2,\dots,N-1\}$.
$W$ is a boolean array of size $N$ with every element initially set to True.
$N_W$ is the number of True elements in array $W$ (initialised to $N$).
$P[i,j]$ is the probability of branching from node $j$ to node $i$.
$AdjIn[i]$ and $AdjOut[i]$ are the lists of indices of all
nodes connected to node $i$ via incoming and outgoing edges, respectively.}

{\bf Recursive function} $F(\alpha,\beta,W,N_W)$
\STATE $W[\beta] \leftarrow$ False; $N_W \leftarrow N_W-1$

\IF{$\alpha=\beta$ and $N_W=0$}
	\STATE $\Sigma \leftarrow 1$
\ELSE
	\STATE $\Sigma \leftarrow 0.0$
	\FORALL{$i \in AdjOut[\beta]$}
		\FORALL{$j \in AdjIn[\beta]$}
			\IF{$W[i]$ and $W[j]$}
				\STATE $\Sigma \leftarrow \Sigma+P[\beta,j] F(j,i,W,N_W) P[i,\beta]$
			\ENDIF
		\ENDFOR
	\ENDFOR
	\STATE $\Sigma \leftarrow 1/(1-\Sigma)$

	\IF{$\alpha \neq \beta$}
		\STATE $\Lambda \leftarrow 0.0$
		\FORALL{$i \in AdjOut[\beta]$}
			\IF{W[i]}
				\STATE $\Lambda \leftarrow \Lambda + F(\alpha,i,W,N_W) P(i,\beta)$
			\ENDIF
		\ENDFOR
		\STATE $\Sigma \leftarrow \Sigma\Lambda$
	\ENDIF
\ENDIF
\STATE $W[\beta] \leftarrow$ True; $N_W \leftarrow N_W+1$
\RETURN $\Sigma$
\end{algorithmic}
\end{algorithm}

\begin{algorithm}
\caption{Calculate the total transition probabilities $\mP^{G_N}_{\alpha,\beta}$ 
and the mean escape times $\mT^{G_N}_{\beta}$ in a dense graph $G_N$}
\label{alg:CGT}
\begin{algorithmic}[1]
\REQUIRE{Nodes are numbered $0,1,2,\dots,N-1$.
Sink nodes are indexed first, source nodes last.
$i$ is the index of the first intermediate node, $s$ is the index of the first source node.
If there are no intermediate nodes then $i=s$, otherwise $i<s$.
$1 < N$ and $i,s \in \{0,1,2,\dots,N-1\}$.
$\tau[\alpha]$ is the waiting time for node $\alpha$, $\alpha\in\{i,i+1,i+2,\dots,N-1\}$.
$P[i,j]$ is the probability of branching from node $j$ to node $i$.}
\FORALL{$\gamma \in \{i,i+1,i+2,\dots,s-1\}$}
\label{line:DetachNodess}
	\FORALL{$\beta \in \{\gamma+1,\gamma+2,\dots,N-1\}$}
		\IF{$P[\gamma,\beta]>0$}
			\STATE $\tau[\beta] \leftarrow (\tau[\beta]+\tau[\gamma] P[\gamma,\beta])/(1-P[\beta,\gamma]P[\gamma,\beta])$
			\FORALL{$\alpha \in \{0,1,2,\dots,N-1\}$}
				\IF{$\alpha\neq\beta$ and $\alpha\neq\gamma$}
					\STATE $P[\alpha,\beta] \leftarrow
					(P[\alpha,\beta]+P[\alpha,\gamma] P[\gamma,\beta])/(1-P[\beta,\gamma]P[\gamma,\beta])$
				\ENDIF
			\ENDFOR
			\STATE $P[\gamma,\beta] \leftarrow 0.0$
		\ENDIF
	\ENDFOR
	\FORALL{$\alpha \in \{0,1,2,\dots,N-1\}$}
		\STATE $P[\alpha,\gamma] \leftarrow 0.0$
	\ENDFOR
\ENDFOR
\label{line:DetachNodesf}
\FORALL{$\alpha \in \{s,s+1,s+2,\dots,N-1\}$}
\label{line:Disconnects}
	\FORALL{$\beta \in \{s,s+1,s+2,\dots,N-1\}$}
		\IF{$\alpha\neq\beta$ and $P[\alpha,\beta]>0$}
			\STATE $P_{\alpha,\beta} \leftarrow P[\alpha,\beta]$; $P_{\beta,\alpha} \leftarrow P[\beta,\alpha]$; $T \leftarrow \tau[\alpha]$
			\STATE $\tau[\alpha] \leftarrow (\tau[\alpha]+\tau[\beta]
			P_{\beta,\alpha})/(1-P_{\alpha,\beta}P_{\beta,\alpha})$
			\STATE $\tau[\beta] \leftarrow (\tau[\beta]+ T P_{\alpha,\beta})/(1-P_{\alpha,\beta}P_{\beta,\alpha})$
			\FORALL{$\gamma \in \{0,1,2,\dots,i-1\}\cup\{s,s+1,s+2,\dots,N-1\}$}
				\STATE $T \leftarrow P[\gamma,\alpha]$
				\STATE $P[\gamma,\alpha] \leftarrow (P[\gamma,\alpha]
				+P[\gamma,\beta]P_{\beta,\alpha})/(1-P_{\alpha,\beta}P_{\beta,\alpha})$
				\STATE $P[\gamma,\beta ] \leftarrow (P[\gamma,\beta ]
				+T P_{\alpha,\beta})/(1-P_{\alpha,\beta}P_{\beta,\alpha})$
			\ENDFOR
			\STATE $P[\alpha,\beta] \leftarrow 0.0$; $P[\beta,\alpha] \leftarrow 0.0$
		\ENDIF
	\ENDFOR
\ENDFOR
\label{line:Disconnectf}
\end{algorithmic}
\end{algorithm}

\begin{algorithm}
\caption{Detach node $\gamma$ from an arbitrary graph $\mathcal{G}_N$}
\label{alg:DetachNode}
\begin{algorithmic}[1]
\REQUIRE{$1 < N$ and $\gamma \in \{0,1,2,\dots,N-1\}$.
$\tau[i]$ is the waiting time for node $i$.
$Adj[i]$ is the ordered list of indices of all nodes connected to node $i$ via outgoing edges.
$|Adj[i]|$ is the cardinality of $Adj[i]$.
$Adj[i][j]$ is the index of the $j$th neighbour of node $i$.
$P[i]$ is the ordered list of probabilities of leaving node $i$ via outgoing edges, $|P[i]|=|Adj[i]|$.
$P[i][j]$ is the probability of branching from node $i$ to node $Adj[i][j]$.}
\FORALL{$\beta_\gamma \in \{0,1,2,\dots,|Adj[\gamma]|-1\}$}
	\STATE $\beta \leftarrow Adj[\gamma][\beta_\gamma]$
	\STATE $\gamma_\beta \leftarrow -1$
	\label{line:findcs}
	\FORALL{$i \in \{0,1,2,\dots,|Adj[\beta]|-1\}$}
		\IF{$Adj[\beta][i]=\gamma$}
			\STATE $\gamma_\beta \leftarrow i$
			\STATE {\bf break}
		\ENDIF
	\ENDFOR
	\label{line:findcf}
	\IF{not $\gamma_\beta=-1$}
		\STATE $P_{\beta,\beta} \leftarrow 1/(1-P[\beta][\gamma_\beta] P[\gamma][\beta_\gamma])$
		\label{line:notsinks}
		\STATE $P_{\beta,\gamma} \leftarrow P[\beta][\gamma_\beta]$
		\STATE $Adj[\beta] \leftarrow
\{Adj[\beta][0],Adj[\beta][1],\dots,Adj[\beta][\gamma_\beta-1],Adj[\beta][\gamma_\beta+1],\dots\}$
		\label{line:dels}
		\STATE $P[\beta] \leftarrow   \{  P[\beta][0],  P[\beta][1],\dots,  P[\beta][\gamma_\beta-1],
P[\beta][\gamma_\beta+1],\dots\}$
		\label{line:delf}
		\FORALL{$\alpha_\gamma \in \{0,1,2,\dots,|Adj[\gamma]|-1\}$}
			\label{line:adds}
			\STATE $\alpha \leftarrow Adj[\gamma][\alpha_\gamma]$
			\IF{not $\alpha=\beta$}
				\IF{exists edge $\beta \rightarrow \alpha$}
					\FORALL{$i \in \{0,1,2,\dots,|Adj[\beta]|-1\}$}
						\label{line:findas}
						\IF{$Adj[\beta][i]=\alpha$}
							\STATE $P[\beta][i] \leftarrow P[\beta][i]+P_{\beta,\gamma}
P[\gamma][\alpha_\gamma]$
							\label{line:pupdate}
							\STATE {\bf break}
						\ENDIF
					\ENDFOR
					\label{line:findaf}
				\ELSE
					\STATE $Adj[\beta] \leftarrow
\{\alpha,Adj[\beta][0],Adj[\beta][1],Adj[\beta][2],\dots\}$
					\label{line:news}
					\STATE $P[\beta] \leftarrow \{P_{\beta,\gamma}
P[\gamma][\alpha_\gamma],P[\beta][0],P[\beta][1],P[\beta][2],\dots\}$
					\label{line:newf}
				\ENDIF
			\ENDIF
		\ENDFOR
		\label{line:addf}
		\FORALL{$i \in \{0,1,2,\dots,|P[\beta]|-1\}$}
		\label{line:renorms}
			\STATE $P[\beta][i] \leftarrow P[\beta][i] P_{\beta,\beta}$
		\ENDFOR
		\label{line:renormf}
		\STATE $\tau_\beta \leftarrow \left( \tau_\beta + P_{\beta,\gamma} \tau_\gamma \right)
P_{\beta,\beta}$
		\label{line:notsinkf}
	\ENDIF
\ENDFOR
\end{algorithmic}
\end{algorithm}

\end{document}